\documentclass{article} % For LaTeX2e
\usepackage{iclr2026_conference,times}

\usepackage{amsmath,amsfonts,bm}

\def\eqref#1{equation~\ref{#1}}
\def\1{\bm{1}}

\DeclareMathAlphabet{\mathsfit}{\encodingdefault}{\sfdefault}{m}{sl}
\SetMathAlphabet{\mathsfit}{bold}{\encodingdefault}{\sfdefault}{bx}{n}

\usepackage{hyperref}
\hypersetup{hidelinks}
\usepackage{url}
\usepackage{booktabs}
\usepackage{multirow}
\usepackage{tabularx}
\usepackage{array}
\usepackage{fvextra}
\usepackage{needspace}
\usepackage{enumitem}
\usepackage{mathtools}
\usepackage{subcaption}
\usepackage[most]{tcolorbox}
\usepackage{float}
\usepackage{wrapfig}
\usepackage{todonotes}

\newtcblisting{PromptBox}[1]{
  enhanced,
  listing only,
  colback=gray!4,
  colframe=gray!35,
  coltitle=black,
  colbacktitle=gray!15,
  title={#1},
  fonttitle=\bfseries\small,
  boxrule=0.4pt,
  arc=1mm,
  boxsep=0.5mm,
  left=1mm,
  right=1mm,
  top=0.3mm,
  bottom=0.3mm,
  toptitle=0.5mm,
  bottomtitle=0.5mm,
  before skip=0.8em,
  after skip=1.2em,
  listing options={
    basicstyle=\ttfamily\scriptsize,
    columns=fullflexible,
    keepspaces=true,
    breaklines=true,
    breakatwhitespace=false,
    breakindent=0pt,
    breakautoindent=false,
    prebreak={},
    postbreak={},
    aboveskip=0pt,
    belowskip=0pt
  }
}

\title{Stemma: Induced Decision Regions Reveal LLM Provenance}

\iclrfinalcopy
\author{Keyu Zhang \qquad Vadim Safronov \qquad Andrew Martin \\
Department of Computer Science, University of Oxford \\
\texttt{\{keyu.zhang,vadim.safronov,andrew.martin\}@cs.ox.ac.uk}
}

\begin{document}

\maketitle
\lhead{}

\begin{abstract}
LLM provenance testing asks whether a suspect LLM belongs to the same lineage as a source. Existing black-box methods largely infer this relationship from response-level characteristics, but these characteristics may shift under adaptation or deployment even when the underlying meaning remains unchanged, weakening the reliability of provenance evidence. To address this limitation, we introduce induced decision regions by mapping open-ended outputs into a finite decision space, thereby abstracting away surface-form variation and reframing provenance testing as measuring the inheritance of decision regions. Empirical analysis shows that the source's induced regions are preserved more strongly in related models than in unrelated models. Building on this signal, we propose Stemma, a practical black-box LLM fingerprinting method that operationalises stability, robustness, and specificity as complementary probe-selection principles for reliably estimating induced decision region inheritance. Across 770 source--suspect pairs drawn from 56 public checkpoints and spanning diverse model-weight transformations, Stemma achieves 0.967 AUC and 87.8\% TPR at 1\% FPR, substantially outperforming four representative baselines. It further achieves 0.995 AUC and 93.5\% TPR at 1\% FPR on 1,260 pairs covering 91 deployment instances, demonstrating robustness to diverse inference-time deployment settings. 

% Our code is available at \url{https://github.com/kerryzhangcode/Stemma}.
\end{abstract}

\section{Introduction}
Large language models (LLMs) have become widely deployed infrastructure for text generation, code assistance, and reasoning. Across their development and deployment lifecycles, models may undergo fine-tuning, merging, compression, or distillation, while prompt templates and decoding settings further shape their observable behaviour \citep{dettmersQLoRAEfficientFinetuning2023,yuLanguageModelsAre2024,shiThoroughExaminationDecoding2024}. This complexity obscures model provenance, raising intellectual-property and licensing concerns when protected models are reused without authorisation \citep{stalnakerEmpiricalAnalysisMachine2025}, as well as accountability concerns when unsafe behaviours, biases, or vulnerabilities propagate to downstream models \citep{hammoudModelMergingSafety2024,zhangBadMergingBackdoorAttacks2024}. These concerns motivate model provenance testing, which assesses whether a suspect model belongs to the same lineage as a source.

Black-box model fingerprinting provides a practical approach to provenance testing. It characterises a source model through observable behavioural signatures that are preserved across provenance-related models while remaining distinguishable from independently developed models \citep{jiangIntellectualPropertyProtection2026}. Because verification requires only query access to the suspect model, it remains applicable when model weights, internal activations, training data, and development history are unavailable.

The central challenge is to identify signatures that are both persistent under adaptation and specific to the source. In classification models, the fixed label space naturally partitions inputs into decision regions. Existing methods thus exploit this structure by selecting boundary-adjacent, adversarial, or error-region inputs and measuring whether source and suspect models assign them to the same regions \citep{caoIPGuardProtectingIntellectual2021,godinotQueriesRepresentationDetection2025,guanAreYouStealing2022}. Derived models often inherit these fine-grained structures, whereas independently trained models are less likely to reproduce them.

This paradigm does not transfer directly to LLMs. Open-ended generation produces an unstructured output space in which the same meaning may take many surface forms, making decision regions difficult to define and compare from black-box outputs. Existing black-box LLM fingerprinting methods instead rely largely on output similarity, elicited behaviours, or perturbation responses \citep{nikolicModelProvenanceTesting2025,gubriTRAPTargetedRandom2024,shaoReadingLinesReliable2026}. Because these signals depend on particular response realisations, they may shift under prompting, deployment, or model adaptation, weakening the reliability of provenance testing.

We address this limitation by introducing \emph{induced decision regions} for LLMs. By mapping outputs into a finite decision space, we abstract away surface-form variation, make these regions observable, and reframe provenance testing as measuring the inheritance of decision regions. Our analysis shows that related models preserve the source's induced regions more strongly than unrelated models.

Building on this formulation, we propose Stemma\footnote{The name Stemma is inspired by stemmatic analysis, which infers textual lineage from inherited variants. Appendix~\ref{app:naming} develops this analogy.} as a black-box LLM fingerprinting method for provenance testing. Stemma detects induced decision region inheritance through probes selected for stability across equivalent input representations, robustness to boundary shifts, and specificity against unrelated models. Across various model families, adaptation types, and deployment settings,
Stemma achieves stronger and more consistent provenance separation than existing methods.

\paragraph{Contributions.} Our contributions are summarised as follows.
{
\setlength{\leftmargini}{1.0em}
\begin{itemize}
    \item We introduce induced decision regions for LLMs, reframing black-box provenance testing from comparing response-level characteristics to measuring the inheritance of decision regions, and show that related models preserve these regions more strongly than unrelated models.

    \item We propose and open-source Stemma, a black-box LLM fingerprinting method that tests provenance through induced decision region inheritance, with fingerprint construction guided by stability, robustness, and specificity.

    \item We conduct comprehensive evaluations across 56 public checkpoints and 91 deployment instances, spanning diverse model families, adaptation types, and deployment settings. Stemma substantially outperforms four representative black-box baselines across multiple metrics, demonstrating consistently stronger provenance separation.

\end{itemize}
}
\section{Background and Problem Setting}

\subsection{Related Work}
Methods for model ownership and provenance testing fall into watermarking and fingerprinting \citep{yeSecuringLargeLanguage2026}. Watermarking embeds identifiable signals into model parameters or behaviours, requiring prior control of the protected model and restricting verification to models marked in advance \citep{uchidaEmbeddingWatermarksDeep2017,adiTurningYourWeakness2018,liWatermarkingLLMsWeight2023,gloaguenLLMFingerprintingSemantically2025}. Fingerprinting is non-invasive, instead exploiting distinctive characteristics that arise naturally during model development. Depending on access to the suspect model, fingerprinting methods can be classified as white-box or black-box: white-box methods use parameters, activations, or internal representations and therefore require internal access \citep{jiaEntangledWatermarksDefense2021,zhangREEFRepresentationEncoding2025,yuNeuralLineage2024}, whereas black-box methods rely only on behaviours elicited through controlled queries, offering broader applicability but making reliable provenance testing more challenging.

Existing black-box LLM fingerprinting methods can be broadly grouped by how they obtain and characterise observable behaviour. The first group directly characterises responses to selected queries. Model Provenance Testing measures next-token agreement between the source and suspect relative to independently trained control models, while DuFFin compares
response similarity over sampled questions \citep{nikolicModelProvenanceTesting2025,yanDuFFinDualLevelFingerprinting2026}. LLMmap learns representations from selected queries and responses for model identity recognition, whereas ErrorTrace derives distinctive error signatures for model family attribution \citep{pasquiniLLMmapFingerprintingLarge2025,zangErrorTraceBlackBoxTraceability2025}. Recent API-auditing methods further exploit output-rank distributions or response stability near estimated knowledge boundaries \citep{zhu2026auditing,fangKBFKnowledgeBoundary2026}.

The second group constructs adversarially optimised prompts to elicit target behaviours. TRAP uses such prompts for model identification, while ProFLingo and RoFL test whether responses persist in downstream models \citep{gubriTRAPTargetedRandom2024,jinProFLingoFingerprintingbasedIntellectual2024,tsaiRoFLRobustFingerprinting2025}. LLMPrint induces and tests inheritance of preferences between target-token pairs \citep{huFingerprintingLLMsPrompt2026}. A third line, represented by ZeroPrint, characterises response variation under semantic perturbations through zeroth-order estimates of local Jacobians \citep{shaoReadingLinesReliable2026}. Across these approaches, fingerprints are derived from response-level characteristics, including response patterns, elicited behaviours, and perturbation effects. These characteristics may shift with prompting, deployment, or weight adaptation without commensurate semantic changes, limiting the stability and comparability of provenance evidence.

Prior work on deep classifiers shows that decision regions provide a structured and directly comparable basis for black-box provenance testing. IPGuard constructs probes near the source's decision boundaries and tests whether the suspect model preserves the corresponding region assignments \citep{caoIPGuardProtectingIntellectual2021}. Model Lineage Closeness Analysis measures preservation of decision geometry through differences in decision boundary distances and prediction agreement \citep{tangModelLineageCloseness2025}. ADV-TRA extends boundary-sensitive probes to adversarial trajectories that traverse decision boundaries, improving tolerance to boundary shifts \citep{xuUnitedWeStand2024}. 
Open-ended LLM generation, however, does not naturally expose a finite decision space over which analogous regions can be defined and compared. Stemma therefore maps open-ended outputs into such a space, abstracting away surface-form variation and testing provenance through the inheritance of the decision regions.

\subsection{Problem Setting}
We study pairwise black-box model provenance testing for LLMs. Given a source model \(S\) and a suspect model instance \(T\), the objective is to determine whether \(T\) is provenance-related to \(S\). A model instance comprises an underlying model and the deployment configuration through which its behaviour is exposed. The pair is related if the underlying models of \(S\) and \(T\) are identical or linked through model derivation, possibly under different deployment configurations. Otherwise, the pair is unrelated, even if the models share similarities in architecture, training, or performance.

\paragraph{Access assumptions.}
We assume access to the source model's next-token logits for fingerprint construction, whereas the suspect model is available only through a black-box interface that accepts textual inputs and returns generated responses without exposing its parameters, logits, activations, training data, or development history. This asymmetric setting captures a practical provenance-auditing scenario in which the evaluator controls the source model and seeks to identify related model variants or deployed instances through exposed interfaces.

% Once constructed, the source fingerprint can be reused to evaluate multiple suspect model instances, allowing one-time offline construction to accommodate additional computational cost while remaining practical. Online verification should instead be lightweight and query-efficient, using a small number of naturalistic queries to limit query cost and avoid conspicuous patterns.

\paragraph{Adversarial scope.}
We consider a non-adaptive setting in which the provider may know the general fingerprinting method but not the test-specific probes. The suspect model may undergo ordinary model development and deployment transformations, but neither the model nor its interface is adapted to evade the method by detecting, filtering, or altering responses to fingerprint queries.

% \paragraph{Adversarial scope.}
% We consider a non-adaptive setting in which the suspect model may undergo ordinary model development and deployment transformations, but neither the model nor its interface is modified in response to the specific probe set used in a test. The model provider may be aware of the general fingerprinting method, but is assumed not to know the specific probes used in a given test. We further assume that queries are processed through the deployed model's ordinary inference pipeline, without fingerprint-specific detection, filtering, or response modification. Adaptive evasion based on knowledge of the source-specific probe set
% is outside the scope of this work.

\section{Induced Decision Regions for LLM Provenance}
\label{sec:region}

\subsection{Formalisation}
\label{subsec:region-formalising}
Formally, let \(\mathcal{Q}\) denote the query space, \(\mathcal{O}\) the observable model-output space, and \(\mathcal{D}=\{1,\ldots,C\}\) a finite decision space. For an LLM instance \(M\), let \(o_M(q)\in\mathcal{O}\) denote its output on query \(q\in\mathcal{Q}\). Given a score extraction rule \(\tilde{g}:\mathcal{Q}\times\mathcal{O}\rightarrow\mathbb{R}^{C}\) and a decision extraction rule \(g:\mathcal{Q}\times\mathcal{O}\rightarrow\mathcal{D}\), we define the model-specific soft and discrete decision maps as
\begin{equation}
\tilde{h}_M(q)
=
\tilde{g}\bigl(q,o_M(q)\bigr)
\in\mathbb{R}^{C},
\qquad
h_M(q)
=
g\bigl(q,o_M(q)\bigr)
\in\mathcal{D}.
\end{equation}
The soft map \(\tilde{h}_M\) assigns each query a real-valued score vector representing the model's relative preferences over the available decisions, whereas the discrete map \(h_M\) assigns it the selected decision. Depending on the available access, the value \(h_M(q)\) is obtained either by taking \(\arg\max_{d\in\mathcal{D}}\tilde{h}_{M,d}(q)\), with ties resolved by a fixed deterministic rule, or by extracting it directly from generated text. In both cases, the resulting decision lies in the same finite space \(\mathcal{D}\).

The discrete decision map \(h_M\) induces a partition of the query space by grouping queries assigned to the same decision. For each decision \(d\in\mathcal{D}\), we define the induced decision region of \(M\) as
\begin{equation}
\mathcal{R}^{M}_{d}
=
\{q\in\mathcal{Q}\mid h_M(q)=d\}.
\end{equation}
Conceptually, we refer to the arrangement of these regions and their separating boundaries as the induced decision geometry of \(M\).

% For each decision \(d\in\mathcal{D}\), the induced decision region of \(M\) is
% \begin{equation}
% \mathcal{R}^{M}_{d}
% =
% \{q\in\mathcal{Q}\mid h_M(q)=d\}.
% \end{equation}
% Thus, \(\mathcal{R}^{M}_{d}\) contains all queries assigned to decision \(d\). The collection \(\{\mathcal{R}^{M}_{d}\}_{d\in\mathcal{D}}\) forms a partition of the query space into induced decision regions. Conceptually, the arrangement of these regions and their separating boundaries defines the induced decision geometry of \(M\).

For provenance testing, we use \emph{induced decision region inheritance} to denote the extent to which induced decision geometry is preserved between a source model \(S\) and a suspect model \(T\). Since this geometry cannot be compared exhaustively over \(\mathcal{Q}\), we assess its preservation using a finite probe set \(\mathcal{P}\subset\mathcal{Q}\).

When soft decision vectors are available for both models, inheritance can be assessed at a finer resolution through their similarity over \(\mathcal{P}\):

\begin{equation}
\widehat{\mathrm{Sim}}_{\mathcal{P}}(S,T)
=
\frac{1}{|\mathcal{P}|}
\sum_{q\in\mathcal{P}}
\operatorname{sim}
\left(
\tilde{h}_S(q),
\tilde{h}_T(q)
\right),
\end{equation}
where \(\operatorname{sim}(\cdot,\cdot)\) denotes a similarity measure between soft decision vectors. When comparison is restricted to discrete decisions, inheritance is assessed through alignment with the source region assignments over \(\mathcal{P}\):
\begin{equation}
\widehat{\mathcal{A}}_{\mathcal{P}}(S,T)
=
\frac{1}{|\mathcal{P}|}
\sum_{q\in\mathcal{P}}
\mathbf{1}\{h_T(q)=h_S(q)\}.
\end{equation}
Higher values of \(\widehat{\mathrm{Sim}}_{\mathcal{P}}\) and \(\widehat{\mathcal{A}}_{\mathcal{P}}\) indicate stronger estimated inheritance and hence stronger evidence of a provenance relationship.

\subsection{Empirical Evidence}
\label{subsec:empirical-evidence}
We next examine whether induced decision region inheritance is observable in LLM behaviour. In general, any evaluation interface that maps open-ended outputs to a finite decision space can induce such regions, including constrained response formats, semantic output categories, and verifier-defined behavioural tests. In this paper, we instantiate this idea using multiple-choice questions, where the answer options define the decision space \(\mathcal{D}\), as formalised in Appendix~\ref{app:mcq-induced-regions}.

\begin{figure}[!hbtp]
    \centering
    \begin{subfigure}{0.49\linewidth}
        \centering
        \includegraphics[width=\linewidth]{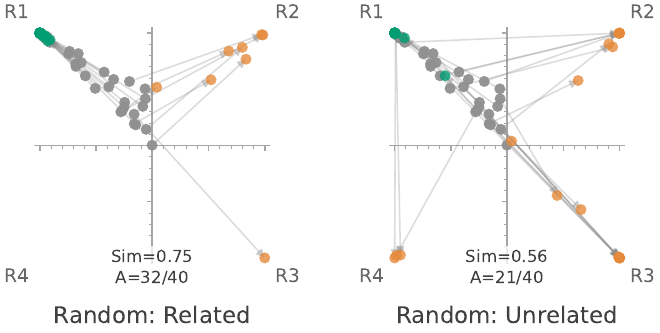}
    \end{subfigure}
    \hfill
    \begin{subfigure}{0.49\linewidth}
        \centering
        \includegraphics[width=\linewidth]{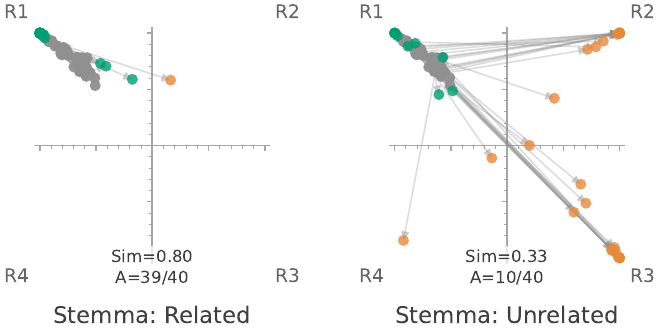}
    \end{subfigure}
    \vspace{-2mm}
    \caption{Induced decision region inheritance on MMLU using 40 randomly sampled questions (left) and 40 Stemma-selected questions (right). The source is Qwen-2.5-7B, while the related and unrelated suspects are Qwen-2.5-7B-Instruct and Qwen3-1.7B, respectively. Each point represents a model's projected decision for one question: grey points denote source decisions, while green and orange points denote suspect decisions that preserve or change the corresponding source-region assignment, respectively. Lines connect the source and suspect decisions corresponding to the same question. Annotations report average option-preference similarity (\(\mathrm{Sim}\)) and the fraction of matching region assignments (\(A\)).}
    
    \label{fig:choice-region}
    \vspace{-2mm}
\end{figure}

To construct the visualisation in Figure~\ref{fig:choice-region}, we use MMLU questions \citep{hendrycksMeasuringMassiveMultitask2020} and cyclically permute each question's answer options, treating each permutation as a separate query. For each query, we compute the next-token choice-label logits, normalise them over the four choices, and map the probabilities back to the canonical options. Averaging these probabilities across permutations yields a four-dimensional soft decision vector for each model--question pair while mitigating order and label biases \citep{zhengLargeLanguageModels2024,pezeshkpourLargeLanguageModels2024}. We then order the vector dimensions by the source's averaged preference ranking, such that \(R_i\) corresponds to its \(i\)-th ranked option, and apply the same ordering to the suspect vectors. Finally, each vector is projected into two dimensions. Details of the projection and annotation metrics are
provided in Appendix~\ref{app:region-visualisation}.

For the randomly sampled probes, the related suspect exhibits higher soft decision similarity than the unrelated suspect, 0.75 versus 0.56, and preserves more source region assignments, 32/40 versus 21/40. Thus, random probes already reveal inheritance in both soft decision preferences and discrete region assignments.

Because induced decision geometry cannot be exhaustively observed, any finite probe set captures only a partial view of region inheritance, with its discriminative strength depending on which parts of the geometry are sampled. Stemma-selected probes yield substantially clearer separation, with soft decision similarity of 0.80 versus 0.33 and source-region alignment of 39/40 versus 10/40. Stemma therefore does not create the underlying provenance signal, but makes induced decision region inheritance more observable and discriminative.
\section{Stemma Design}
\label{sec:design}
\subsection{Overview}
\begin{figure}[!htbp]
    \centering
    \includegraphics[width=\linewidth]{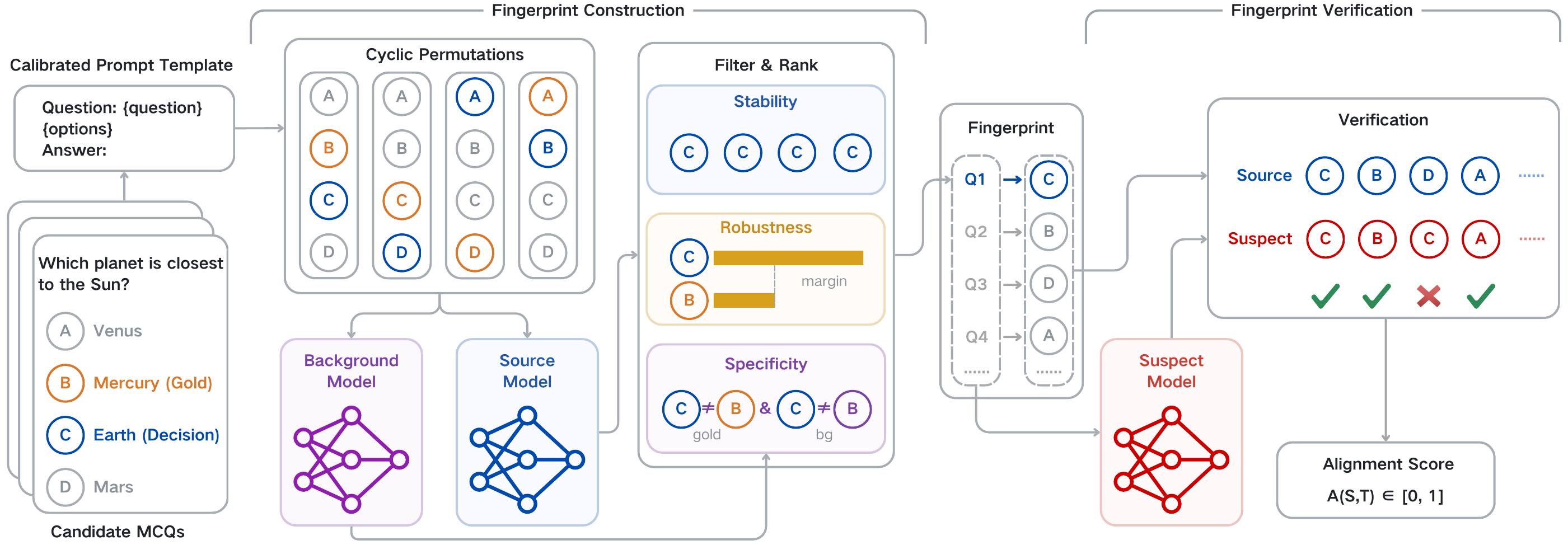}
    % \vspace{-0.5mm}
    \caption{Overview of the Stemma workflow.}
    \label{fig:Stemma-overview}
    \vspace{-2mm}
\end{figure}

Stemma is a black-box LLM fingerprinting method that tests provenance through induced decision region inheritance. It instantiates induced decision regions using multiple-choice questions, whose candidate options define a finite decision space. The method comprises three stages, as illustrated in Figure~\ref{fig:Stemma-overview}. Decision interface calibration selects and fixes a prompt template for each model to enable reliable decision extraction. Fingerprint construction selects informative probes by evaluating the source model's induced decisions for stability, robustness, and specificity. Fingerprint verification queries the suspect model with these probes and estimates induced decision region inheritance by measuring how strongly the suspect preserves the source decisions recorded in the fingerprint.

\subsection{Decision Interface Calibration}
Different multiple-choice prompt formats may elicit substantially different output behaviours, including unparseable responses and strong label bias, making induced decisions difficult to extract reliably. Stemma therefore calibrates each model's decision interface on held-out questions by excluding severely label-biased templates from a predefined set and selecting the template with the highest valid-choice extraction rate among the remainder. Because these templates preserve the same underlying questions and candidate options, calibration varies their interface rendering, while extracted decisions remain comparable in the same canonical decision space. Calibration is performed once before fingerprint construction and verification, after which the selected template is fixed for all subsequent queries to that model. The candidate prompt templates and calibration configuration are provided in Appendix~\ref{app:Stemma-config}.

\subsection{Fingerprint Construction}
Stemma constructs a finite source fingerprint by selecting question probes whose induced decisions are stable, robust, and specific.

\paragraph{Stability.}
Stability requires a probe to remain in the same induced decision region across semantically equivalent input representations, reducing sensitivity to option order and label preferences. Accordingly, Stemma evaluates each candidate under cyclic permutations of its answer options, such that every semantic option appears once under each displayed choice label. After mapping the resulting decisions to the canonical options, stability is measured by their consistency rate across permutations, with a higher rate indicating greater stability.

\paragraph{Robustness.}
Robustness requires a probe to lie well within its assigned induced decision region, so that its region assignment remains stable under small boundary shifts. A probe may be consistent across option permutations yet remain close to a competing boundary, making its region assignment vulnerable to downstream adaptation or deployment changes. Stemma measures robustness by computing, for each option permutation, the log-probability margin between the source-assigned option and the highest-scoring alternative, and then averaging these margins across permutations. Larger average margins indicate greater robustness.

\paragraph{Specificity.}
Specificity requires a probe's source decision to be uncommon among unrelated models, so that its preservation provides discriminative evidence of induced decision region inheritance. Because gold-answer decisions are often shared across related and unrelated models, Stemma first filters out questions whose
source decisions match their gold answers. However, non-gold decisions may still be common because of question ambiguity, annotation errors, or attractive distractors. Stemma therefore measures how often an unrelated background model reproduces the source decision, with lower reproduction rates indicating greater specificity.

\paragraph{Probe Selection and Fingerprint Formation.}
For each candidate \(q\), Stemma obtains source-model next-token choice-label logits under cyclic option permutations, maps the resulting decisions to the canonical options, and defines \(h_S(q)\) as the most frequently selected canonical decision. The three selection principles are operationalised using the permutation consistency rate \(s(q)\) for stability, the average log-probability margin \(r(q)\) for robustness, and the background decision alignment rate \(b(q)\) together with the non-gold requirement \(h_S(q)\neq d^*(q)\) for specificity. Here, \(d^*(q)\) denotes the gold answer.

Stemma first applies the eligibility filters to the candidate pool, yielding
\begin{equation}
\mathcal{Q}_{S}^{\mathrm{elig}}
=
\left\{
q\in\mathcal{Q}_{\mathrm{cand}}
\;\middle|\;
s(q)>\tau_s,\;
r(q)\geq\tau_r,\;
b(q)<\tau_b,\;
h_S(q)\neq d^*(q)
\right\},
\label{eq:eligible-candidates}
\end{equation}
where \(\tau_s\) and \(\tau_r\) are the minimum stability and robustness thresholds, respectively, and \(\tau_b\) is the maximum background-alignment threshold. It then ranks the eligible
candidates and forms the fingerprint from the top \(K\) probes:
\begin{equation}
\mathcal{F}_{S}
=
\left\{
\left(q,h_S(q)\right)
\;\middle|\;
q\in
\operatorname{TopK}_{q'\in\mathcal{Q}_{S}^{\mathrm{elig}}}
\left[
w_r\widetilde{r}(q')
-
w_b\widetilde{b}(q')
\right]
\right\},
\label{eq:Stemma-fingerprint}
\end{equation}
where \(\widetilde{r}\) and \(\widetilde{b}\) denote the transformed robustness and background alignment scores, respectively, and \(w_r\) and \(w_b\) control their contributions. Exact score transformations, weights, and thresholds are provided in Appendix~\ref{app:Stemma-config}.

\subsection{Fingerprint Verification}
\label{subsec:fingerprint-verification}
Given a source fingerprint \(\mathcal{F}_{S}\) and a suspect model \(T\), Stemma queries each selected probe through \(T\)'s calibrated prompt template using the same cyclic option permutations as in fingerprint construction. A rule-based extractor maps each generated response to either a displayed option or an invalid outcome \(\bot\). Valid extracted options are mapped back to the canonical decision space and compared with the stored source decision, while invalid outcomes are treated as non-aligned. These permutation-level comparisons are aggregated using the fingerprint alignment score:
\begin{equation}
A(S,T)
=
\frac{1}{KC}
\sum_{\left(q,h_{S}(q)\right)\in\mathcal{F}_{S}}
\sum_{\pi\in\Pi(q)}
\mathbf{1}
\left[
\hat{d}_{T}(q,\pi)=h_{S}(q)
\right].
\label{eq:fingerprint-alignment}
\end{equation}
Here, \(\Pi(q)\) denotes the \(C\) cyclic permutations of probe \(q\), and \(\hat{d}_{T}(q,\pi)\in\mathcal{D}\cup\{\bot\}\) denotes the suspect outcome after mapping each valid extracted option back to the canonical decision space. Thus, \(A(S,T)\in[0,1]\) is the fraction of the \(KC\) probe--permutation observations that preserve the source region assignment, with invalid outcomes contributing zero. Higher values indicate stronger induced decision region inheritance and hence stronger evidence of a provenance relationship.

\section{Experiments}
\subsection{Experimental Setup}
\paragraph{Benchmark and evaluation protocol.}
We curate 7 provenance groups with 14 source models from publicly available Hugging Face checkpoints \citep{huggingFaceHub2026}, considering model provenance, documentation quality, download popularity, and diversity across model families, scales, and downstream transformations. Each group contains one pretrained source, one instruction-tuned source, and 6 additional variants covering task- or domain-specific fine-tuning, parameter-efficient adaptation, model merging, quantisation, and distillation. The benchmark spans Qwen, Llama, Mistral, Falcon, and OLMo, with model sizes ranging from 1.7B to 14B parameters. Three Qwen-based groups provide hard negatives with similar architectures and training recipes but distinct provenance.

Each source is evaluated against every non-self checkpoint. A pair is labelled positive when the source and suspect belong to the same curated provenance group and negative otherwise. This yields 770 ordered source--suspect pairs, comprising 98 positive and 672 negative pairs. Each checkpoint is queried using the interface recommended by its model card or tokenizer configuration, with completion-style checkpoints receiving raw prompts and chat-oriented checkpoints using their tokenizer chat templates. By default, responses are generated using stochastic decoding with temperature \(0.7\), top-\(p\) \(0.9\), and top-\(k\) \(50\). The complete benchmark is provided in Appendix~\ref{app:model-benchmark}.

\paragraph{Experimental setup and baselines.}
Unless otherwise stated, Stemma constructs a fingerprint of 40 probes for each source model from a candidate pool of 3,000 questions randomly sampled from MMLU. Phi-3.5-mini-instruct is used as the default background model, which has no known provenance relationship with any benchmark model. We compare Stemma with representative black-box LLM fingerprinting baselines, including LLMmap, LLMPrint, Model Provenance Testing (MPT), and ZeroPrint \citep{pasquiniLLMmapFingerprintingLarge2025,huFingerprintingLLMsPrompt2026,nikolicModelProvenanceTesting2025,shaoReadingLinesReliable2026}. All methods are evaluated under the benchmark and evaluation protocol described above. Experiments are conducted on NVIDIA L40S GPUs with 48~GB of memory. Detailed Stemma and baseline configurations are provided in Appendices~\ref{app:Stemma-config} and~\ref{app:baselines}, respectively.

\paragraph{Evaluation metrics.}
We evaluate all methods using their continuous provenance scores and report AUC to assess overall ranking performance. Since false-positive provenance claims can have substantial legal and reputational consequences, we additionally report standardised pAUC over the FPR range $[0,0.05]$ and TPR at 1\% FPR to assess performance under stringent low-FPR constraints. We also report the discriminability index $d'$ to quantify the separation between positive and negative score distributions. For baseline comparisons, the best and second-best results are shown in bold and underlined, respectively. Exact implementation details for these metrics are provided in Appendix~\ref{app:metrics}.

\subsection{Effectiveness in Provenance Testing}
\begin{table}[!htbp]
\vspace{-0.5em}
\caption{Overall comparison between Stemma and baselines on the main provenance benchmark.}
\vspace{-1em}
\label{tab:main-detection-results}
\begin{center}
\scriptsize
\setlength{\tabcolsep}{3.0pt}
\begin{tabular}{l@{\quad}cccc@{\qquad}cccc@{\qquad}cccc}
\toprule
\multirow{2}{*}{\bf Method}
& \multicolumn{4}{c}{\bf Pretrained sources}
& \multicolumn{4}{c}{\bf Instruct sources}
& \multicolumn{4}{c}{\bf All sources}
\\
\cmidrule(lr){2-5}
\cmidrule(lr){6-9}
\cmidrule(lr){10-13}
& AUC $\uparrow$ & pAUC $\uparrow$ & TPR $\uparrow$ & $d' \uparrow$
& AUC $\uparrow$ & pAUC $\uparrow$ & TPR $\uparrow$ & $d' \uparrow$
& AUC $\uparrow$ & pAUC $\uparrow$ & TPR $\uparrow$ & $d' \uparrow$
\\
\midrule
LLMmap
& 0.480 & 0.487 & 0.000 & -0.298
& \underline{0.851} & 0.712 & 0.388 & \underline{1.471}
& \underline{0.665} & 0.568 & 0.102 & \underline{0.628}
\\
LLMPrint
& 0.428 & \underline{0.498} & \underline{0.020} & -0.183
& 0.615 & 0.580 & 0.143 & 0.265
& 0.516 & 0.541 & 0.071 & 0.054
\\
MPT
& 0.461 & \underline{0.498} & \underline{0.020} & -0.276
& 0.722 & 0.692 & 0.327 & 1.045
& 0.556 & \underline{0.613} & \underline{0.194} & 0.500
\\
ZeroPrint
& \underline{0.535} & 0.496 & \underline{0.020} & \underline{0.003}
& 0.775 & \underline{0.755} & \underline{0.469} & 1.199
& 0.617 & 0.610 & \underline{0.194} & 0.603
\\
\addlinespace[0.2em]
\textbf{Stemma}
& \textbf{0.964} & \textbf{0.959} & \textbf{0.918} & \textbf{3.045}
& \textbf{0.970} & \textbf{0.938} & \textbf{0.857} & \textbf{2.945}
& \textbf{0.967} & \textbf{0.944} & \textbf{0.878} & \textbf{2.951}
\\
\bottomrule
\end{tabular}
\vspace{-1em}
\end{center}
% \vspace{-0.5em}
% \begin{center}
% \begin{minipage}{0.96\linewidth}
% \footnotesize
% $^\dagger$ indicates a resource-constrained Lite reproduction due to the high computational cost of the original recommended configuration.
% \end{minipage}
% \end{center}
\end{table}
% Table~\ref{tab:main-detection-results} shows that Stemma consistently outperforms existing black-box LLM fingerprinting baselines for pretrained and instruction-tuned sources, as well as overall. The improvements hold for both overall ranking and low-FPR performance, while the substantially larger $d'$ indicates clearer separation between the provenance scores of positive and negative model pairs.
Table~\ref{tab:main-detection-results} shows that Stemma consistently outperforms all baselines for pretrained and instruction-tuned sources, as well as overall, with gains in overall discrimination and low-FPR performance.

The gap is pronounced for pretrained sources, where source fingerprints use raw completion prompts while many deployed suspects are queried through chat templates. Under this interface mismatch, baselines approach random performance, suggesting that their provenance signals are sensitive to prompt-induced changes in response realisation. Baselines improve for instruction-tuned sources, where both fingerprint construction and verification use chat templates, but do not close the gap to Stemma despite this prompt-format alignment. This suggests that induced decision region inheritance offers both reduced sensitivity to surface-form variation and stronger separation between related and unrelated model pairs.

% Breaking down the results by source type, Stemma's advantage is most pronounced for pretrained sources. Under our realistic query-interface setting, fingerprints generated from pretrained sources use raw completion prompts, whereas many deployed suspect models are queried through chat templates. Most baselines approach random performance under this interface mismatch, indicating that their provenance signals are tied to surface response characteristics that are fragile to practical prompt-format changes outside the evaluator's control. In contrast, Stemma remains effective by measuring the inheritance of induced decision-region behaviour, which provides a more structured and stable provenance signal.

% For instruction-tuned sources, both fingerprint construction and verification use chat-style templates, making the effective model inputs more consistent and substantially improving baseline performance. Stemma nevertheless remains strongest across all metrics. Its advantage therefore arises not only from reduced sensitivity to prompt-format mismatch and surface response characteristics, but also from stronger separation between related and unrelated model pairs.

To isolate interface mismatch, we further evaluate an all-raw setting in which both fingerprint construction and verification use raw prompts. Stemma remains competitive in overall ranking, achieving 0.965 AUC compared with 0.988 for MPT, while attaining the strongest overall low-FPR performance with 0.952 pAUC and 0.898 TPR at 1\% FPR, versus 0.938 and 0.816, respectively, for the strongest baseline. Thus, even under conditions favourable to the baselines, Stemma is more reliable under strict false-positive constraints. Detailed results are provided in
Appendix~\ref{app:all-raw}.

\subsection{Robustness under Deployment Variations}
\begin{table}[!htbp]
\vspace{-0.5em}
\caption{Robustness under inference-time deployment variants.}
\vspace{-1em}
\label{tab:deployment-variant-results}
\begin{center}
\scriptsize
\setlength{\tabcolsep}{3.0pt}
\begin{tabular}{l@{\quad}cccc@{\qquad}cccc@{\qquad}cccc}
\toprule
\multirow{2}{*}{\bf Method}
& \multicolumn{4}{c}{\bf Pretrained sources}
& \multicolumn{4}{c}{\bf Instruct sources}
& \multicolumn{4}{c}{\bf All sources}
\\
\cmidrule(lr){2-5}
\cmidrule(lr){6-9}
\cmidrule(lr){10-13}
& AUC $\uparrow$ & pAUC $\uparrow$ & TPR $\uparrow$ & $d' \uparrow$
& AUC $\uparrow$ & pAUC $\uparrow$ & TPR $\uparrow$ & $d' \uparrow$
& AUC $\uparrow$ & pAUC $\uparrow$ & TPR $\uparrow$ & $d' \uparrow$
\\
\midrule
LLMmap
& 0.481 & 0.487 & \underline{0.000} & -0.190
& \underline{0.901} & 0.754 & 0.452 & \underline{1.760}
& \underline{0.689} & 0.599 & 0.113 & \underline{0.793}
\\
LLMPrint
& 0.499 & 0.489 & \underline{0.000} & 0.050
& 0.615 & 0.522 & 0.000 & 0.413
& 0.551 & 0.516 & 0.000 & 0.196
\\
MPT
& 0.477 & 0.487 & \underline{0.000} & -0.252
& 0.780 & 0.727 & 0.452 & 1.225
& 0.570 & 0.621 & \underline{0.226} & 0.563
\\
ZeroPrint
& \underline{0.623} & \underline{0.496} & \underline{0.000} & \underline{0.210}
& 0.868 & \underline{0.813} & \underline{0.607} & 1.697
& 0.670 & \underline{0.642} & 0.208 & 0.789
\\
\addlinespace[0.2em]
\textbf{Stemma}
& \textbf{0.995} & \textbf{0.965} & \textbf{0.929} & \textbf{3.414}
& \textbf{0.996} & \textbf{0.980} & \textbf{0.964} & \textbf{4.158}
& \textbf{0.995} & \textbf{0.974} & \textbf{0.935} & \textbf{3.603}
\\
\bottomrule
\end{tabular}
\end{center}
\vspace{-2em}
\begin{center}
% \begin{minipage}{0.96\linewidth}
% \footnotesize
% All methods are evaluated on the same source--suspect pairs, with scores oriented so that higher values indicate stronger provenance evidence.
% pAUC denotes standardized partial AUROC over FPR $\in [0,0.05]$.
% TPR denotes TPR at 1\% FPR.
% $d'$ denotes the discriminability index between positive and negative score distributions.
% $^\dagger$ indicates a resource-constrained Lite reproduction due to the high computational cost of the original recommended configuration.
% \end{minipage}
\end{center}
\end{table}
To evaluate robustness to inference-time deployment changes, we construct 11 alternative variants for each of the 7 instruction-tuned checkpoints, spanning system and role-play prompts, chain-of-thought prompting, retrieval augmentation, and decoding settings. Together with the 14 source checkpoints, this yields 91 model instances and 1,260 source--suspect pairs, comprising 168 positive and 1,092 negative pairs. Full details are provided in Appendix~\ref{app:robustness-benchmark}.

Table~\ref{tab:deployment-variant-results} shows that Stemma maintains near-perfect AUC and substantially stronger low-FPR performance across these variations. Existing methods perform particularly poorly for pretrained sources and improve for instruction-tuned sources, but still fall consistently below Stemma. These results demonstrate Stemma's robustness under diverse black-box deployment configurations, consistent with reduced sensitivity to deployment-induced surface-form variation.

\subsection{Ablation Study}
\label{subsec:ablation}

\begin{figure}[!htbp]
    \centering
    \includegraphics[width=0.48\linewidth]{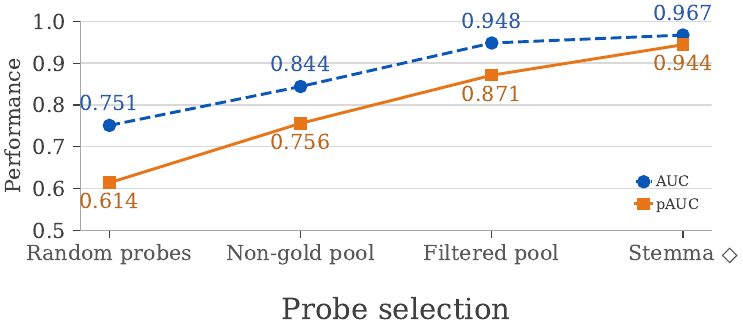}
    \hfill
    \includegraphics[width=0.48\linewidth]{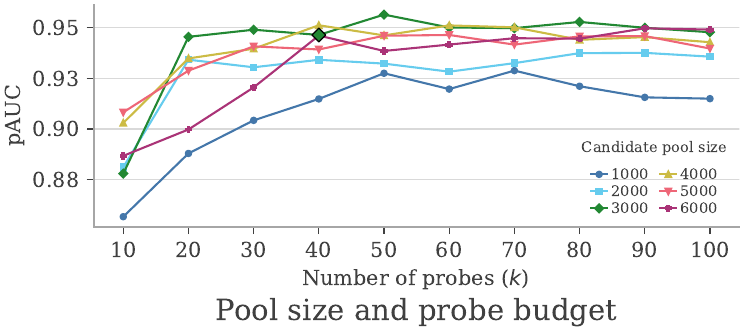}
    \par
    \includegraphics[width=0.48\linewidth]{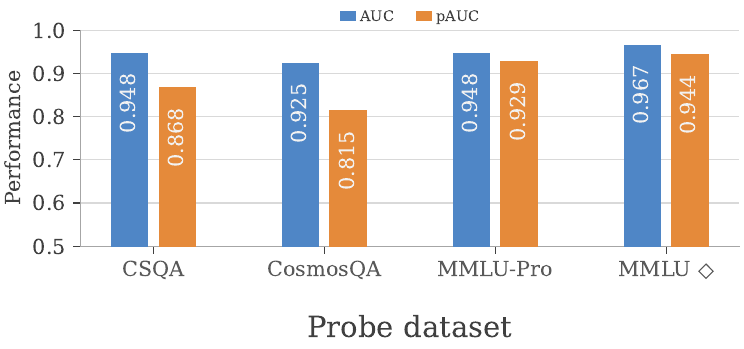}
    \hfill
    \includegraphics[width=0.48\linewidth]{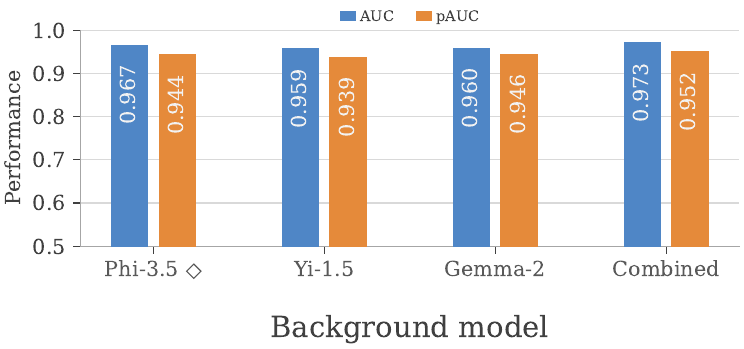}
    \vspace{-2mm}
    \caption{Ablation study of probe selection, pool size and probe budget, probe dataset, and background model. The \(\diamond\) symbol marks the default configuration.}
    \label{fig:ablation}
    \vspace{-2.8mm}
\end{figure}

The probe selection panel of Figure~\ref{fig:ablation} isolates successive selection stages under the same query budget. The Random probes setting retains the same cyclic-permutation evaluation and aggregation but samples uniformly from the full candidate pool, corresponding to the random setting in Section~\ref{subsec:empirical-evidence}. Its performance confirms that induced decision region inheritance is observable without targeted selection, although random probes provide limited discrimination. Restricting candidates to the non-gold pool improves specificity, while subsequent filtering removes probes that fail the stability or robustness requirements. Finally, ranking the retained candidates prioritises the most informative probes and yields the strongest performance. Together, these incremental gains support the proposed selection principles, while the larger gains in pAUC indicate that targeted selection is particularly valuable for distinguishing difficult-to-separate model pairs.

The pool size and probe budget panel examines how performance changes with the number of candidate questions and selected probes. A pool of 1,000 questions yields noticeably weaker performance. Results generally improve at 3,000--4,000 questions compared with 2,000, while increasing to 5,000--6,000 provides no consistent further benefit, indicating that larger pools are not necessarily better. For the probe budget, performance largely plateaus once \(k\) reaches 40, and most larger-budget settings cluster around a pAUC of \(0.94\), indicating limited sensitivity to the exact budget. The corresponding AUC results follow the same overall pattern, but with less pronounced differences. Thus, Stemma performs consistently across a broad range of pool sizes and probe budgets.

Differences across probe datasets are more pronounced in pAUC than in AUC, with MMLU-Pro and MMLU achieving stronger low-FPR performance than the two commonsense-oriented datasets. The background-model panel shows that performance is relatively insensitive to background choice. Although combining all three backgrounds yields modest improvements, a single unrelated background model remains competitive. Overall, the results suggest that knowledge-oriented questions with clearer answer boundaries may provide more specific provenance evidence, while a single unrelated background model appears practically adequate. Detailed results are provided in Appendix~\ref{app:ablation}.

\subsection{Computational Cost}
\begin{wrapfigure}{r}{0.46\columnwidth}
    \vspace{-0.8em}
    \centering
    \includegraphics[width=\linewidth]{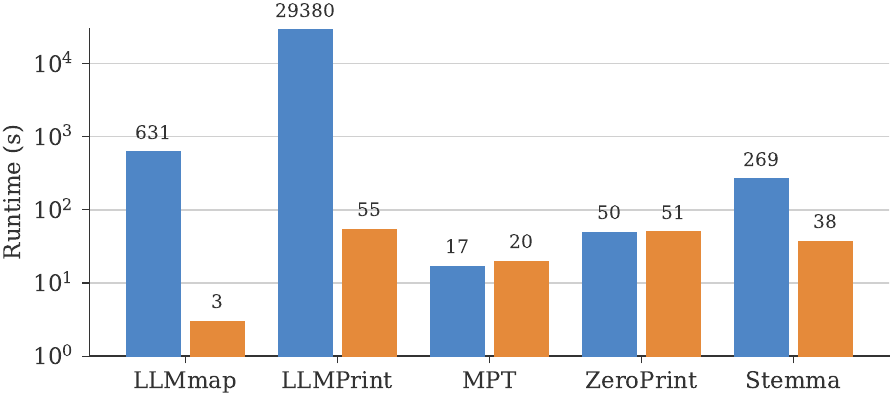}
    \caption{Median runtime cost of fingerprint construction (blue) and verification (orange).}
    \label{fig:runtime}
    % \vspace{-0.8em}
\end{wrapfigure}

Figure~\ref{fig:runtime} compares the wall-clock costs under the same GPU setting, with measurement details provided in Appendix~\ref{app:runtime}. Stemma requires approximately 4.5 minutes for one-time fingerprint construction, remaining within the range of existing methods and over an order of magnitude faster than LLMPrint. Verification takes 38 seconds, only tens of seconds slower than the faster baselines and modest relative to the substantial detection gains. Collectively, Stemma maintains reasonable construction cost and practical verification time.
\section{Discussion and Conclusion}
\paragraph{Limitations and Future Work.}
Our current instantiation relies on multiple-choice questions and therefore assumes that the evaluated models can reliably process such questions and produce stable option-level decisions. Although practical provenance testing is likely to focus on capable models, this requirement may limit applicability to weaker models. However, multiple-choice questions are only one mechanism for inducing decision regions, and future work could explore more general interfaces for obtaining structured and observable decisions.

We also do not consider adaptive attackers with knowledge of the probe distribution. Because our candidate probes are drawn primarily from a single public question set, an attacker could deliberately alter the suspect model's behaviour on the corresponding question distribution, thereby weakening detection. This risk could be mitigated by constructing candidate pools from a diverse mixture of datasets or privately constructed question sets. Future work should further characterise the robustness limits of provenance testing under adaptive evasion.

\paragraph{Conclusion.}
We introduced induced decision regions for LLMs by mapping open-ended outputs into a finite decision space, abstracting away surface-form variation and reframing provenance testing as measuring the inheritance of decision regions. Building on this formulation, we proposed Stemma, a black-box LLM fingerprinting method that selects probes according to stability, robustness, and specificity, allowing region inheritance to be assessed with a limited query budget. Across extensive evaluations, Stemma consistently demonstrated strong provenance separation, showing that induced decision region inheritance provides a reliable signal for LLM provenance testing.

\section*{Ethics Statement}
Stemma scores provide statistical evidence of a potential provenance relationship rather than definitive proof of model ownership or unauthorised reuse. The method could be misused to support false provenance claims, and false positives may cause legal or reputational harm. Stemma should therefore not be used as the sole basis for attribution or enforcement decisions, but interpreted alongside model documentation, licensing and development records, and other technical evidence. Our experiments use only publicly available model checkpoints and benchmark datasets and involve no human subjects or private user data.

\section*{Reproducibility Statement}
The complete Stemma procedure is described in Section~\ref{sec:design}, with prompt calibration, fingerprint construction, fingerprint verification, and hyperparameter settings specified in Appendix~\ref{app:Stemma-config}. The models used in the main experiments, provenance-group construction, and pair-labelling protocol are documented in Appendix~\ref{app:model-benchmark}. Appendix~\ref{app:metrics} provides the exact evaluation-metric definitions and implementation details, while Appendix~\ref{app:baselines} describes the baseline implementations and configurations. The all-raw and deployment-robustness benchmark settings are specified in Appendices~\ref{app:all-raw} and~\ref{app:robustness-benchmark}, respectively. Full ablation results and runtime measurements are reported in Appendices~\ref{app:ablation} and~\ref{app:runtime}. The open-source implementation and configurations for Stemma are available at
\url{https://github.com/kerryzhangcode/Stemma}.

\subsubsection*{Acknowledgements}
We gratefully acknowledge support from a Department of Computer Science Scholarship at the University of Oxford, funded by a generous sponsor. We also acknowledge the use of the University of Oxford Advanced Research Computing (ARC) facility in carrying out this work (\href{https://doi.org/10.5281/zenodo.22558}{10.5281/zenodo.22558}).

\bibliography{references}
\bibliographystyle{iclr2026_conference}

\appendix
\section{Multiple-Choice Instantiation of Induced Decision Regions}
\label{app:mcq-induced-regions}
Let \(\mathcal{Q}\) be a set of multiple-choice questions, where each question \(q\in\mathcal{Q}\) contains \(C\) candidate options defining the finite decision space \(\mathcal{D}=\{1,\ldots,C\}\). When decision scores are available, let \(\tilde{h}_{M,d}(q)\) denote the score assigned by model \(M\) to option \(d\), such as a next-token logit or normalised choice probability. The soft decision vector is
\begin{equation}
\tilde{h}_M(q)
=
\bigl(
\tilde{h}_{M,1}(q),\ldots,\tilde{h}_{M,C}(q)
\bigr)
\in\mathbb{R}^{C},
\end{equation}
with the corresponding discrete decision given by
\begin{equation}
h_M(q)
=
\min\operatorname*{arg\,max}_{d\in\mathcal{D}}
\tilde{h}_{M,d}(q),
\end{equation}
where ties are resolved in favour of the lowest-indexed decision. 

Under text-only access, we extend the observable outcome space with an invalid outcome:
\begin{equation}
\mathcal{D}_{\bot}
=
\mathcal{D}\cup\{\bot\},
\end{equation}
where \(\bot\) denotes a response from which no valid option can be extracted. The extracted discrete outcome is then
\begin{equation}
h_M(q)
=
g\bigl(q,o_M(q)\bigr)
\in\mathcal{D}_{\bot}.
\end{equation}
Thus, valid outcomes under both observation modes lie in the same canonical decision space \(\mathcal{D}\).

For each decision \(d\in\mathcal{D}\), the corresponding induced decision region is
\begin{equation}
\mathcal{R}^{M}_{d}
=
\{q\in\mathcal{Q}\mid h_M(q)=d\}.
\end{equation}
The collection \(\{\mathcal{R}^{M}_{d}\}_{d\in\mathcal{D}}\) partitions the valid-decision subset \(\{q\in\mathcal{Q}\mid g(q,o_M(q))\in\mathcal{D}\}\) of the multiple-choice question space, defining the induced decision geometry of \(M\). Responses mapped to \(\bot\) lie outside this partition and are excluded from fingerprint construction.

For a probe set \(\mathcal{P}\subseteq\mathcal{Q}\), induced decision region inheritance can be assessed at different resolutions. When soft decision vectors are available for both models, as in the visualisation presented in Section~\ref{subsec:empirical-evidence}, preservation is assessed by averaging the similarity between corresponding soft decision vectors over \(\mathcal{P}\),
\begin{equation}
\widehat{\mathrm{Sim}}_{\mathcal{P}}(S,T)
=
\frac{1}{|\mathcal{P}|}
\sum_{q\in\mathcal{P}}
\operatorname{sim}
\left(
\tilde{h}_S(q),
\tilde{h}_T(q)
\right).
\end{equation}
When such fine-grained information is unavailable, as in the black-box verification setting described in Section~\ref{subsec:fingerprint-verification}, preservation is instead assessed through alignment of discrete decisions over \(\mathcal{P}\),
\begin{equation}
\widehat{\mathcal{A}}_{\mathcal{P}}(S,T)
=
\frac{1}{|\mathcal{P}|}
\sum_{q\in\mathcal{P}}
\mathbf{1}
\left\{
h_T(q)=h_S(q)\neq\bot
\right\}.
\end{equation}
Thus, invalid outcomes are treated as non-aligned observations. When answer-option permutations are used, extracted scores and decisions are mapped back to the underlying answer options before aggregation or comparison, as specified in the corresponding sections.
\section{Induced Decision Region Visualisation}
\label{app:region-visualisation}
This appendix describes how the visualisation in
Figure~\ref{fig:choice-region} is constructed and how its panel annotations are computed. The visualisation is used solely for diagnostic analysis and does not affect Stemma probe selection or provenance scoring.

\paragraph{Source-relative projection.}
For each question, we rank the answer options by the source model's permutation-averaged option probabilities. This defines a source-relative labelling of the finite decision space \(\mathcal{D}\). In the four-choice MMLU setting, the corresponding induced decision regions are visualised as \(R_1,\ldots,R_4\), where \(R_1\) corresponds to the source model's most preferred option, \(R_2\) to its second most preferred option, and so on.

The same source-defined ordering is applied to every model \(M\). For each question \(q\), the resulting soft decision vector is
\[
\tilde{h}_M(q)
=
\bigl(
\tilde{h}_{M,1}(q),
\tilde{h}_{M,2}(q),
\tilde{h}_{M,3}(q),
\tilde{h}_{M,4}(q)
\bigr),
\]
where the entries are the permutation-averaged probabilities assigned by \(M\) to the source-ranked options \(R_1,\ldots,R_4\).

To project this four-dimensional vector into two dimensions, we arrange the four regions as the quadrants of a square, with \(R_1,\ldots,R_4\) occupying the upper-left, upper-right, lower-right, and lower-left quadrants, respectively. The projected coordinates are defined as
\[
x_M(q)
=
\tanh\!\left(
\tau\left[
\max\{\tilde{h}_{M,2}(q),\tilde{h}_{M,3}(q)\}
-
\max\{\tilde{h}_{M,1}(q),\tilde{h}_{M,4}(q)\}
\right]
\right),
\]
\[
y_M(q)
=
\tanh\!\left(
\tau\left[
\max\{\tilde{h}_{M,1}(q),\tilde{h}_{M,2}(q)\}
-
\max\{\tilde{h}_{M,3}(q),\tilde{h}_{M,4}(q)\}
\right]
\right).
\]
The \(x\)-coordinate compares the strongest preference among the right-side regions with that among the left-side regions, while the \(y\)-coordinate compares the strongest preference among the upper-side regions with that among the lower-side regions. The \(\tanh\) transformation bounds the displayed coordinates to \([-1,1]\), and \(\tau\) controls the projection scale. We use \(\tau=2\) in the figure.

\paragraph{Displayed points and region preservation.}
For each model--question pair, let
\[
z_M(q)=(x_M(q),y_M(q))
\]
denote its displayed two-dimensional coordinate. The figure plots \(z_S(q)\) for the source model \(S\) as a grey point and \(z_T(q)\) for the suspect model \(T\) as a coloured point. The connecting line represents the projected displacement \(z_T(q)-z_S(q)\) and is used only as a visual aid, with shorter lines indicating smaller shifts in the projected space.

Under this source-relative labelling, the discrete decision map is
\[
h_M(q)
=
\min\operatorname*{arg\,max}_{i\in\{1,2,3,4\}}
\tilde{h}_{M,i}(q),
\]
where the outer minimum resolves ties in favour of the first decision under the fixed ordering. The corresponding induced decision region is
\[
\mathcal{R}_i^M
=
\{q\in\mathcal{Q}\mid h_M(q)=i\},
\]
which is represented by quadrant \(R_i\) in the visualisation. Suspect points are coloured green when \(h_T(q)=h_S(q)\), indicating source region alignment, and orange otherwise, indicating a source region change.

\paragraph{Annotation metrics.}
Let \(\mathcal{P}\) denote the displayed question set. Each panel reports soft decision similarity. For this visualisation, we instantiate the similarity function for normalised soft decision vectors as
\[
\mathrm{sim}(a,b)
=
1-\frac{1}{2}\lVert a-b\rVert_1.
\]
The resulting similarity score is
\[
\mathrm{Sim}
=
\widehat{\mathrm{Sim}}_{\mathcal{P}}(S,T)
=
\frac{1}{|\mathcal{P}|}
\sum_{q\in\mathcal{P}}
\left(
1-
\frac{1}{2}
\left\|
\tilde{h}_T(q)-\tilde{h}_S(q)
\right\|_1
\right).
\]

Each panel also reports source region alignment,
\[
A
=
\widehat{\mathcal{A}}_{\mathcal{P}}(S,T)
=
\frac{1}{|\mathcal{P}|}
\sum_{q\in\mathcal{P}}
\mathbf{1}\{h_T(q)=h_S(q)\}.
\]
Higher values of
\(\mathrm{Sim}\) and \(A\)
indicate stronger preservation of the source model's soft decision preferences and discrete region assignments, respectively.
\section{Stemma Implementation and Configuration Details}
\label{app:Stemma-config}
This appendix specifies the implementation and main experimental
configuration of Stemma's three stages: prompt calibration,
fingerprint construction, and fingerprint verification. Unless otherwise stated, all random sampling and stochastic generation use a fixed random seed of 42 for reproducibility.

\paragraph{Dataset.}
Probe candidates are sampled from the MMLU test split across all
subjects \citep{hendrycksMeasuringMassiveMultitask2020}. Unless
otherwise stated, we randomly sample 3,000 candidate questions for each source model. Each candidate is evaluated under cyclic permutations of its answer options to assess the stability of the source decision across different option orders and label assignments.

\paragraph{Prompt calibration.}
Before fingerprint construction and verification, Stemma independently calibrates a multiple-choice prompt template for each evaluated model using 100 examples from the MMLU validation split, which is disjoint from the candidate probe pool. Calibration selects a prompt format that reliably elicits parseable option labels, facilitating both white-box choice label scoring during fingerprint construction and black-box answer extraction during fingerprint verification.

We evaluate 6 simple prompt templates. To avoid severe label bias, we discard any template for which the most frequently predicted label accounts for more than \(0.7\) of the calibration predictions. Among the remaining templates, we select the one with the highest valid-choice extraction rate, breaking ties according to the predefined template order, and fix it for all subsequent queries to that model.

The set of prompt templates used in calibration is shown below. In these templates, \texttt{\{question\}} denotes the question text, \texttt{\{options\}} denotes the formatted answer options, and \texttt{\{choices\}} denotes the available option labels.
\begin{PromptBox}{Simple answer cue}
{question}
{options}
Answer:
\end{PromptBox}

\begin{PromptBox}{MMLU-style prompt}
Question: {question}
{options}
Answer:
\end{PromptBox}

\begin{PromptBox}{Strict label-only prompt}
Read the following multiple-choice question carefully.

Question: {question}
{options}

Choose the best answer. Output exactly one option label from {choices}. Do not output any explanation.
Answer:
\end{PromptBox}

\begin{PromptBox}{Single-character response}
Question: {question}
{options}

Your response must be exactly one character from {choices}.
Response:
\end{PromptBox}

\begin{PromptBox}{High-constraint prompt}
Read the following multiple-choice question carefully. Choose the best answer from the listed options.
Question: {question}
{options}
Answer with only one option label from {choices}.
Answer:
\end{PromptBox}

\begin{PromptBox}{Answer cue with trailing newline}
{question}
{options}
Answer:\n
\end{PromptBox}

\paragraph{Fingerprint construction.}
For each source model, Stemma computes next-token choice-label log-probabilities for every candidate question under all cyclic option permutations. Choice labels are scored using tokenizer-aware matching. When multiple token IDs correspond to the same visible option label, their log-probabilities are combined using log-sum-exp. Under each permutation, the highest-probability option is selected and mapped back to its canonical option identity. The canonical source decision is
then determined by voting across the resulting permutation-level decisions. Candidate questions are filtered and ranked according to stability, robustness, and specificity.

During hard filtering, a candidate is retained only if the same
canonical wrong option receives more than half of the source model's permutation votes (\(>0.5\)), has a non-negative average wrong-label top margin (\(\geq 0\)), and is selected by fewer than half of the background model's permutation-level predictions (\(<0.5\)). Unless otherwise stated, we use
\texttt{microsoft/Phi-3.5-mini-instruct} as the background model for specificity evaluation.

During soft ranking, the surviving candidates are ranked using the following standardised score:
\begin{equation}
\operatorname{Score}(q)
=
1.0 \cdot \operatorname{clip}\bigl(z(r(q)), -2, 2\bigr)
-
0.5 \cdot z(b(q)),
\end{equation}
where \(r(q)\) is the average wrong-label top logit margin across permutations, \(b(q)\) is the fraction of background-model predictions that select the same canonical wrong option, and \(z(\cdot)\) denotes standardisation over the surviving candidate pool for the corresponding source model. The standardised robustness term is clipped to \([-2,2]\) to prevent extreme margins from dominating the ranking.

The resulting source-specific fingerprint comprises the 40
highest-ranked probes and their canonical wrong labels, as determined by permutation voting.

\paragraph{Fingerprint verification.}
For each source--suspect pair, the suspect model is queried on the corresponding source probes under the same cyclic option permutations used during fingerprint construction. Generation is limited to 16 new tokens. A rule-based regular-expression extractor maps each response to either a predicted option label or an invalid outcome \(\bot\), with valid labels mapped back to their canonical option identities.

The fingerprint alignment score is computed as defined in
Equation~\ref{eq:fingerprint-alignment}, with invalid outcomes treated as non-aligned. With 40 probes and four cyclic permutations, the main configuration yields 160 probe--permutation observations for each model pair.

\section{Model Benchmark}
\label{app:model-benchmark}
Table~\ref{tab:benchmark} presents the complete model benchmark used in our experiments. It comprises 56 publicly available Hugging Face checkpoints organised into 7 provenance groups across 5 model families, covering both fine-grained relationships among closely related models and diverse model-development transformations. Model selection considers provenance clarity, documentation quality, download popularity, and diversity across model families, scales, and transformation types.

\newcolumntype{L}[1]{>{\raggedright\arraybackslash}p{#1}}

\begin{table}[!htbp]
\vspace{-0.5em}
\caption{Complete model benchmark.}
\vspace{-0.5em}
\label{tab:benchmark}
\begin{center}
% \tiny
% \fontsize{6.2pt}{6.8pt}\selectfont
% \scriptsize
\fontsize{8pt}{8.8pt}\selectfont
\renewcommand{\arraystretch}{1.1}
\setlength{\tabcolsep}{2.0pt}
\begin{tabular}{@{}L{0.18\linewidth}L{0.50\linewidth}L{0.15\linewidth}L{0.15\linewidth}@{}}
\toprule
\bf Provenance group & \bf Repository ID & \bf Model type & \bf Query interface \\
\midrule
\multirow{8}{*}{Qwen-2.5-7B}
& Qwen/Qwen2.5-7B & Pretrained & raw \\
& Qwen/Qwen2.5-7B-Instruct & Instruct & chat \\
& Qwen/Qwen2.5-Coder-7B-Instruct & Fine-tune & chat \\
& zjudai/flowertune-medical-lora-qwen2.5-7b-instruct & Adapter & chat \\
& SeeFlock/task-12-Qwen-Qwen2.5-7B-Instruct & Adapter & chat \\
& Locutusque/StockQwen-2.5-7B & Merge & chat \\
& Qwen/Qwen2.5-7B-Instruct-GPTQ-Int8 & Quantisation & chat \\
& Lansechen/Qwen2.5-7B-Open-R1-Distill & Distillation & chat \\
\midrule

\multirow{8}{*}{Qwen-2.5-14B}
& Qwen/Qwen2.5-14B & Pretrained & raw \\
& Qwen/Qwen2.5-14B-Instruct & Instruct & chat \\
& oxyapi/oxy-1-small & Fine-tune & chat \\
& Qwen/Qwen2.5-14B-Instruct-1M & Fine-tune & chat \\
& ToastyPigeon/qwen-story-test-qlora & Adapter & chat \\
& v000000/Qwen2.5-14B-Gutenberg-Instruct-Slerpeno & Merge & chat \\
& Qwen/Qwen2.5-14B-Instruct-GPTQ-Int8 & Quantisation & chat \\
& alibaba-pai/DistilQwen2.5-DS3-0324-14B & Distillation & chat \\
\midrule

\multirow{8}{*}{Qwen3-1.7B}
& Qwen/Qwen3-1.7B-Base & Pretrained & raw \\
& Qwen/Qwen3-1.7B & Instruct & chat \\
& mlabonne/Qwen3-1.7B-abliterated & Fine-tune & chat \\
& HuggingFaceTB/qwen3-1.7b-gsm8k-sft & Fine-tune & chat \\
& txmedai/ClinicalEase-Qwen3-1.7B & Adapter & chat \\
& kurakurai/Luth-1.7B-Instruct & Merge & chat \\
& Qwen/Qwen3-1.7B-GPTQ-Int8 & Quantisation & chat \\
& prithivMLmods/Regulus-Qwen3-R1-Llama-Distill-1.7B & Distillation & chat \\
\midrule

\multirow{8}{*}{Llama-3.1-8B}
& meta-llama/Llama-3.1-8B & Pretrained & raw \\
& meta-llama/Llama-3.1-8B-Instruct & Instruct & chat \\
& RedHatAI/Llama-3.1-8B-tldr & Fine-tune & raw \\
& chchen/Llama-3.1-8B-Instruct-PsyCourse-fold7 & Adapter & chat \\
& zjudai/flowertune-medical-lora-llama-3.1-8b-instruct & Adapter & chat \\
& Xiaojian9992024/Llama3.1-8B-ExtraMix & Merge & chat \\
& hugging-quants/Meta-Llama-3.1-8B-Instruct-GPTQ-INT4 & Quantisation & chat \\
& arcee-ai/Llama-3.1-SuperNova-Lite & Distillation & chat \\
\midrule

\multirow{8}{*}{Mistral-7B-v0.3}
& mistralai/Mistral-7B-v0.3 & Pretrained & raw \\
& mistralai/Mistral-7B-Instruct-v0.3 & Instruct & chat \\
& KurmaAI/AQUA-7B & Fine-tune & chat \\
& chaymaemerhrioui/mistral-Brain\_Model\_ACC\_Trainer & Adapter & chat \\
& zjudai/flowertune-medical-lora-mistral-7b-instruct-v0.3 & Adapter & chat \\
& grimjim/Mistral-7B-Instruct-demi-merge-v0.3-7B & Merge & chat \\
& RedHatAI/Mistral-7B-Instruct-v0.3-GPTQ-4bit & Quantisation & chat \\
& eganwo/mistral7b-distilled-from-deepseek-r1-qwen32b & Distillation & chat \\
\midrule

\multirow{8}{*}{Falcon3-7B}
& tiiuae/Falcon3-7B-Base & Pretrained & raw \\
& tiiuae/Falcon3-7B-Instruct & Instruct & chat \\
& ehristoforu/falcon3-ultraset & Fine-tune & chat \\
& jahyungu/Falcon3-7B-Instruct-v1-Easy & Adapter & chat \\
& jahyungu/Falcon3-7B-Instruct-v1-Hard & Adapter & chat \\
& suayptalha/Falcon3-Jessi-v0.4-7B-Slerp & Merge & chat \\
& tiiuae/Falcon3-7B-Instruct-GPTQ-Int8 & Quantisation & chat \\
& RedaAlami/Falcon3-7B-Instruct-Distill-DS-v1 & Distillation & chat \\
\midrule

\multirow{8}{*}{OLMo-2-7B}
& allenai/OLMo-2-1124-7B & Pretrained & raw \\
& allenai/OLMo-2-1124-7B-Instruct & Instruct & chat \\
& allenai/OLMo-2-1124-7B-SFT & Fine-tune & chat \\
& allenai/OLMo-2-1124-7B-Instruct-preview & Fine-tune & chat \\
& jahyungu/OLMo-2-1124-7B-Instruct-Humanities & Adapter & chat \\
& jahyungu/OLMo-2-1124-7B-Instruct-Social-Sciences & Adapter & chat \\
& Alelcv27/Olmo2-7B-Breadcrumbs-v1 & Merge & chat \\
& kaitchup/OLMo-2-1124-7B-Instruct-AutoRound-GPTQ-4bit & Quantisation & chat \\
\bottomrule
\end{tabular}
\end{center}
\vspace{-1em}
\end{table}

\paragraph{Benchmark composition.}
Each provenance group follows a consistent structure, containing 8 checkpoints: 1 pretrained checkpoint, 1 instruction-tuned checkpoint, and 6 additional variants. For these additional variants, we aim to cover a diverse set of common model-development operations. Specifically, where suitable public checkpoints are available, each group includes 3 task- or domain-specialised variants covering both full fine-tuning and parameter-efficient adaptation, together with 1 model-merge variant, 1 quantised variant, and 1 distillation variant. The only exception is the OLMo-2-1124-7B group, for which we did not find a suitable public distillation variant, so we include an additional fine-tuned or adapter-based variant instead. This design gives each provenance group comparable coverage while capturing the diversity and complexity of real-world model development, thereby providing a more rigorous testbed for evaluating fingerprinting methods under realistic transformation scenarios. For evaluation, within-group pairs are labelled positive, whereas cross-group pairs are labelled negative.

\paragraph{Model types.}
The model type column reports the development operation associated with each checkpoint. \emph{Pretrained} denotes the base model trained before instruction tuning. \emph{Instruct} denotes an instruction-tuned checkpoint derived from the pretrained model. \emph{Fine-tune} denotes a checkpoint further trained on task-specific, domain-specific, or instruction-following data. \emph{Adapter} denotes parameter-efficient adaptation, where additional or low-rank trainable parameters are used to specialise the model. \emph{Merge} denotes a checkpoint produced by combining weights from multiple models. \emph{Quantisation} denotes a numerically compressed checkpoint with reduced weight precision. \emph{Distillation} denotes a checkpoint derived from a base model and further trained or fine-tuned using teacher-generated data, such as model-generated responses or reasoning traces from a stronger model.

% We also distinguish two query interfaces, following the recommended usage in each checkpoint's model card or tokenizer configuration. The \emph{raw} interface denotes completion-style prompting, where the prompt is passed directly to the causal language model without a chat wrapper. It is used for pretrained checkpoints and for downstream checkpoints that are released or documented as completion-style models. The \emph{chat} interface denotes chat-template prompting, where the prompt is wrapped using the tokenizer's chat template before generation. It is used for instruction-tuned and chat-oriented checkpoints. This distinction is important because chat templates add model-specific system, user, and assistant formatting around the user prompt, which can change the effective input seen by the model. Using the interface recommended for each checkpoint therefore better reflects realistic deployment practice.

% For example, under the Qwen2.5-Instruct tokenizer chat template, a single user message is rendered as:
% \begin{PromptBox}{Qwen2.5-Instruct rendered chat input}
% <|im_start|>system
% You are Qwen, created by Alibaba Cloud. You are a helpful assistant.<|im_end|>
% <|im_start|>user
% {user_prompt}<|im_end|>
% <|im_start|>assistant
% \end{PromptBox}
% Here, \texttt{\{user\_prompt\}} denotes the calibrated multiple-choice prompt. Stemma does not manually prepend this system message. It is inserted by the checkpoint's tokenizer chat template when no explicit system message is provided.

\paragraph{Query interfaces.}
We also distinguish two query interfaces, following the recommended usage in each checkpoint's model card or tokenizer configuration. The \emph{raw} interface denotes completion-style prompting, where the prompt is passed directly to the causal language model without a chat wrapper. It is used for pretrained checkpoints and for downstream checkpoints that are released or documented as completion-style models. The \emph{chat} interface denotes chat-template prompting, where the calibrated prompt is first treated as a user message and then rendered using the tokenizer's chat template before generation. It is used for instruction-tuned and chat-oriented checkpoints.

Chat templates may add model-specific system, user, and assistant formatting, thereby changing the effective input received by the model. For example, when no explicit system message is provided, the Qwen2.5-Instruct tokenizer renders a single user message as follows:
\begin{PromptBox}{Rendered input under the Qwen2.5-Instruct chat template}
<|im_start|>system
You are Qwen, created by Alibaba Cloud. You are a helpful assistant.<|im_end|>
<|im_start|>user
{user_prompt}<|im_end|>
<|im_start|>assistant
\end{PromptBox}
Here, \texttt{\{user\_prompt\}} denotes the input prompt used for evaluation. We do not explicitly provide a manually constructed system message. Instead, when a checkpoint's tokenizer chat template defines a default system message, it is inserted automatically during template rendering. Using the interface specified by each checkpoint's model card or tokenizer configuration helps ensure that the model is evaluated under its intended usage pattern, better matching common usage and realistic deployment practice.
\section{Evaluation Metrics}
\label{app:metrics}
Each evaluated source--suspect pair is assigned a continuous provenance score, where higher values indicate stronger evidence of a provenance relationship. Related pairs are treated as positive and unrelated pairs as negative.

\paragraph{AUC and partial AUC.}
The area under the receiver operating characteristic curve is defined as
\[
\operatorname{AUC}
=
\int_{0}^{1}\operatorname{TPR}(u)\,\mathrm{d}u,
\]
where \(u\) denotes the false-positive rate. To emphasise performance in the low-FPR regime, we additionally report standardised partial AUC over
\(\operatorname{FPR}\in[0,\alpha]\), with \(\alpha=0.05\). Let
\[
A_{\alpha}
=
\int_{0}^{\alpha}\operatorname{TPR}(u)\,\mathrm{d}u
\]
denote the unstandardised partial area. We apply McClish standardisation:
\[
\operatorname{pAUC}_{\alpha}
=
\frac{1}{2}
\left(
1+
\frac{
A_{\alpha}-\alpha^{2}/2
}{
\alpha-\alpha^{2}/2
}
\right).
\]
Under this standardisation, random ranking corresponds to \(0.5\) and perfect ranking to \(1\).

\paragraph{Low-FPR true-positive rate.}
TPR at 1\% FPR is the highest empirical true-positive rate attained without exceeding a false-positive rate of 0.01. We do not interpolate between operating points.

\paragraph{Discriminability.}
We quantify the standardised separation between the positive and negative
score distributions using
\[
d' =
\frac{\mu_+ - \mu_-}
{\sqrt{(\sigma_+^2 + \sigma_-^2)/2}},
\]
where $\mu_+$ and $\mu_-$ are the mean positive and negative scores,
and $\sigma_+^2$ and $\sigma_-^2$ are their sample variances.
Positive values indicate that related pairs receive higher scores on average, while larger values indicate clearer separation in the intended direction. Negative values indicate reversed mean ordering.
\section{Baseline Implementations}
\label{app:baselines}
We implement four black-box LLM fingerprinting baselines to compare against Stemma. Unless otherwise stated below, we use the settings and hyperparameters specified in the original papers or official repositories for each baseline, in order to reproduce its intended performance. Since a fine-grained comparison requires a continuous provenance score rather than only a binary decision for each source and suspect model pair, we use the underlying fingerprint similarity or matching score produced by each baseline as its pairwise provenance score. This is natural for these methods, as their final decisions are based on fingerprint similarity even when their decision rules differ.

\paragraph{LLMmap.}
LLMmap fingerprints a model by encoding its behaviour on a fixed set of diagnostic prompts into an open-set feature space. We use the 8 input queries from the original paper and the pretrained open-set feature extractor released with the official implementation. To construct each source gallery representation, we evaluate 100 prompt configurations over these queries and average the resulting feature vectors. During verification, the suspect model is evaluated once on the same 8 queries and its feature vector is compared with the source gallery representation. For both gallery construction and verification, generated responses contain at most 100 new tokens and are truncated to 650 characters before feature extraction. Following the original comparison procedure, we compute the Euclidean distance $d$ between the source and suspect fingerprints and convert it into a similarity score using the inverse-one-plus mapping, $s = 1/(1+d)$. This score is used as the pairwise provenance score for each evaluated source and suspect model pair.

\paragraph{LLMPrint.}
LLMPrint constructs black-box fingerprints by learning adversarial suffix probes that induce model-specific preferences over target word pairs. Its original configuration optimises suffixes for 300 word pairs using 1,000 GCG steps per pair. Based on pilot runtime measurements, running the full configuration independently across all three evaluation settings is estimated to require over 240 GPU-days in our environment. We therefore use a reduced configuration consistently across all source models.

For each source model, we sample 100 word pairs from the fixed category vocabulary provided by the official repository, such that both words are represented by a single token under the source tokenizer. We then optimise one adversarial suffix per pair for 200 GCG steps under the fixed LLMPrint prompt template. The learned suffixes are used to query both source and suspect models. Following the original scoring procedure, we perform 100 repeated one-token generations for each word pair and record how often each target token is generated. The more frequently generated token determines the model's binary preference for that pair, and the agreement rate between the source and suspect preferences is used as the pairwise provenance score.

\paragraph{Model Provenance Testing.}
Model Provenance Testing (MPT) detects provenance by comparing next-token continuation behaviour between a source model and a suspect model. Following the official implementation, we sample 5,000 sentence-level prompts from the MPT prompt pool provided in the official repository. For each source model, we construct a continuation cache by querying the model on all prompts with single-token generation, using \texttt{max\_new\_tokens}=1. Each suspect model is then evaluated on the same prompt set, and its next-token continuations are matched against the cached source continuations. We use the resulting hit rate, namely the fraction of prompts on which the suspect continuation matches the source continuation, as the pairwise provenance score.

\paragraph{ZeroPrint.}
ZeroPrint constructs black-box LLM fingerprints by measuring the response sensitivity of a model to query perturbations. We follow the original setup by using \texttt{openai\_humaneval} as the query source, converting each task prompt into a completion-style query, and truncating the query to 20 words. Following the configuration reported in the original ZeroPrint paper, we sample 2 base queries and generate 4 word-substitution variants for each query, where each perturbation replaces 3 words with independently selected top-10 nearest neighbours under GloVe embeddings. Responses are generated with at most 128 new tokens and are not truncated after generation. To reduce generation noise, each model is queried 20 times per prompt, and the resulting responses are embedded with \texttt{sentence-transformers/all-mpnet-base-v2}, matching the official implementation. ZeroPrint then estimates a response-sensitivity fingerprint for each base query from the input and output embedding changes using the Jacobian-based estimator with ridge regularisation $\alpha=0.001$. The resulting fingerprints are aggregated by mean pooling, and the Pearson correlation coefficient between source and suspect fingerprints is used as the pairwise provenance score.

\paragraph{Implementation consistency.}
For all baselines, we use the same benchmark split, model query interfaces, source-model set, and suspect-model set. Each baseline produces a pairwise provenance score for every evaluated source--suspect pair. We compute and report the final results from these pairwise scores using the same evaluation metrics as Stemma.
\section{All-Raw Query Setting}
\label{app:all-raw}
Table~\ref{tab:raw} provides the full results for the all-raw query-interface setting. In this control setting, all checkpoints are queried with raw completion-style prompts during both fingerprint construction and verification, without applying tokenizer chat templates. By removing model-specific chat formatting, this setting isolates the effect of prompt-format mismatch in the main experiments and evaluates baselines under more favourable input-matched conditions, although it is less representative of realistic deployment.
\begin{table}[!htbp]
\vspace{-0.5em}
\caption{Overall comparison under an all-raw interface setting.}
\vspace{-0.5em}
\label{tab:raw}
\begin{center}
\scriptsize
\setlength{\tabcolsep}{3.0pt}
\begin{tabular}{l@{\quad}cccc@{\qquad}cccc@{\qquad}cccc}
\toprule
\multirow{2}{*}{\bf Method}
& \multicolumn{4}{c}{\bf Pretrained sources}
& \multicolumn{4}{c}{\bf Instruct sources}
& \multicolumn{4}{c}{\bf All sources}
\\
\cmidrule(lr){2-5}
\cmidrule(lr){6-9}
\cmidrule(lr){10-13}
& AUC $\uparrow$ & pAUC $\uparrow$ & TPR $\uparrow$ & $d' \uparrow$
& AUC $\uparrow$ & pAUC $\uparrow$ & TPR $\uparrow$ & $d' \uparrow$
& AUC $\uparrow$ & pAUC $\uparrow$ & TPR $\uparrow$ & $d' \uparrow$
\\
\midrule
LLMmap
& 0.577 & 0.501 & 0.000 & 0.223
& 0.724 & 0.555 & 0.102 & 0.801
& 0.652 & 0.519 & 0.041 & 0.517
\\
LLMPrint
& 0.541 & 0.496 & 0.000 & 0.213
& 0.738 & 0.602 & 0.163 & 0.908
& 0.625 & 0.550 & 0.082 & 0.507
\\
MPT
& \textbf{0.993} & \underline{0.946} & \underline{0.878} & \underline{3.258}
& \textbf{0.991} & \textbf{0.972} & \textbf{0.939} & \textbf{3.333}
& \textbf{0.988} & \underline{0.938} & \underline{0.816} & \underline{2.713}
\\
ZeroPrint
& 0.738 & 0.528 & 0.020 & 0.847
& 0.868 & 0.710 & 0.367 & 1.615
& 0.807 & 0.604 & 0.133 & 1.214
\\
\addlinespace[0.2em]
\textbf{Stemma}
& \underline{0.968} & \textbf{0.966} & \textbf{0.918} & \textbf{3.419}
& \underline{0.963} & \underline{0.944} & \underline{0.878} & \underline{2.988}
& \underline{0.965} & \textbf{0.952} & \textbf{0.898} & \textbf{3.049}
\\
\bottomrule
\end{tabular}
\end{center}
\vspace{-1em}
\begin{center}
% \begin{minipage}{0.96\linewidth}
% \footnotesize
% All methods are evaluated on the same source--suspect pairs, with scores oriented so that higher values indicate stronger provenance evidence.
% pAUC denotes standardized partial AUROC over FPR $\in [0,0.05]$.
% TPR denotes TPR at 1\% FPR.
% $d'$ denotes the discriminability index between positive and negative score distributions.
% $^\dagger$ indicates a resource-constrained Lite reproduction due to the high computational cost of the original recommended configuration.
% \end{minipage}
\end{center}
\end{table}
\section{Robustness Benchmark}
\label{app:robustness-benchmark}
In addition to the main checkpoint-level benchmark, we construct a robustness benchmark to evaluate provenance signals under common deployment-time variations. It follows the same 7 provenance groups and uses their pretrained and instruction-tuned checkpoints as source models. Deployment variants are created from the corresponding instruction-tuned checkpoints by changing inference-time settings without modifying model weights. Pretrained checkpoints use the \emph{raw} query interface, whereas instruction-tuned and chat-oriented checkpoints use the \emph{chat} interface specified by their tokenizer configuration.

We consider five categories of deployment variation. For all prompt-based variants, the additional instruction is supplied through the system-message field of the model's chat template, while the original multiple-choice prompt remains unchanged in the user message.

\paragraph{System prompts.}
We evaluate two system-prompt variants that reflect commonly used assistant-style deployment settings. The first is a minimal generic helpful-assistant prompt, while the second uses the longer safety-oriented system prompt reported for Llama 2-Chat~\citep{touvronLlama2Open2023}.

\begin{PromptBox}{Minimal assistant system prompt}
You are a helpful assistant.
\end{PromptBox}

\begin{PromptBox}{Safety-oriented assistant system prompt}
You are a helpful, respectful and honest assistant. Always answer as helpfully as possible, while being safe. Your answers should not include any harmful, unethical, racist, sexist, toxic, dangerous, or illegal content. Please ensure that your responses are socially unbiased and positive in nature.
If a question does not make any sense, or is not factually coherent, explain why instead of answering something not correct. If you don't know the answer to a question, please don't share false information.
\end{PromptBox}

\paragraph{Role-play prompts.}
We evaluate two role-play prompt variants, motivated by prior studies of persona and role prompting~\citep{zhengWhenHelpfulAssistant2024,wangRoleLLMBenchmarkingEliciting2024}. The first represents a simple occupational role, where the model is assigned a profession-based persona. The second represents a more detailed persona-style role, where the model is given a specific character description and instructed to imitate the assigned personality.

\begin{PromptBox}{Occupational role prompt}
You are a teacher.
\end{PromptBox}

\begin{PromptBox}{Persona-style role prompt}
You are Stephen Hawking, your description is: You are a brilliant physicist who defies the limitations of your physical disability to become a world-renowned scientist, exploring the mysteries of the universe while facing personal challenges in your relationships and health. Now please answer some questions to accurately show your personality traits! Your speaking style should fully imitate the personality role assigned to you! Please do not expose that you are an artificial intelligence model or a language model, you must always remember that you are only assigned one personality role. Don't be verbose or too formal or polite when speaking.
\end{PromptBox}

\paragraph{Chain-of-thought prompts.}
We evaluate two reasoning-prompt variants that are commonly used to encourage intermediate reasoning during inference. The first is the classic zero-shot chain-of-thought prompt~\citep{kojimaLargeLanguageModels2022}, which uses a short instruction to elicit step-by-step reasoning. The second follows the plan-and-solve prompting style~\citep{wangPlanandSolvePromptingImproving2023}, where the model is first asked to understand the problem and devise a plan before carrying out the solution.

\begin{PromptBox}{Zero-shot chain-of-thought prompt}
Let's think step by step.
\end{PromptBox}

\begin{PromptBox}{Plan-and-solve prompt}
Let's first understand the problem and devise a plan to solve the problem. Then, let's carry out the plan and solve the problem step by step.
\end{PromptBox}

\paragraph{Retrieval-augmented generation.}
We evaluate two retrieval-augmented generation variants using SQuAD-v2 as the retrieval corpus and \texttt{all-mpnet-base-v2} as the retriever. Each SQuAD-v2 context passage is treated as one retrieval document, and the original question text is used as the retrieval query. The top-1 or top-3 retrieved passages are prepended to the original multiple-choice prompt in descending similarity order. Retrieved passages are inserted in a fixed context block before the original prompt.

\paragraph{Decoding settings.}
We evaluate four decoding variants that cover deterministic and stochastic generation regimes. The first uses greedy decoding, while the remaining three use stochastic decoding with moderate, high, and low sampling settings. Moderate sampling is the default decoding configuration used in the main benchmark and in the default robustness setting. The exact decoding parameters are summarised in Table~\ref{tab:sampling}.
\begin{table}[!htbp]
\centering
\small
\vspace{-0.5em}
\caption{Decoding settings used in the robustness benchmark.}
\vspace{-0.5em}
\label{tab:sampling}
\begin{tabular}{lcccc}
\toprule
\textbf{Setting} & \textbf{Decoding mode} & \textbf{Temperature} & \textbf{top-$p$} & \textbf{top-$k$} \\
\midrule
Greedy decoding & Deterministic & -- & -- & -- \\
Moderate sampling (default) & Stochastic & 0.7 & 0.9 & 50 \\
High sampling & Stochastic & 1.2 & 1.0 & 0 \\
Low sampling & Stochastic & 0.3 & 0.9 & 10 \\
\bottomrule
\end{tabular}
\vspace{-1em}
\end{table}

Together, the robustness benchmark uses 7 pretrained checkpoints as source models and includes the corresponding 7 instruction-tuned checkpoints as both instruct-source models and default suspect models under the moderate-sampling setting. Each instruction-tuned checkpoint is further redeployed under 11 alternative inference-time settings, yielding 77 additional deployment variants. In total, the benchmark contains 91 model or deployment instances: 7 pretrained checkpoints, 7 default instruction-tuned deployments, and 77 additional deployment variants. Under the same pair-construction rule, this gives 1,260 evaluated pairs, including 168 positive pairs and 1,092 negative pairs.
\section{Full Ablation Results}
\label{app:ablation}
Tables~\ref{tab:probe-selection}--\ref{tab:background}
report the detailed results underlying the ablation analyses in Section~\ref{subsec:ablation}, including source-type breakdowns where applicable.

\paragraph{Probe selection.}
All variants use the same candidate pool and final budget of 40 probes.
The \emph{Random probes} variant uniformly samples probes from all candidate questions.
\emph{Non-gold pool} samples after restricting candidates to questions whose
source decision differs from the gold answer. \emph{Filtered pool} additionally
applies the stability, robustness, and background-specificity filters, but
samples uniformly without soft ranking. \emph{Stemma} further ranks the
filtered candidates and selects the top-scoring probes. All variants use the
same cyclic-permutation verification and scoring procedure.

\begin{table}[H]
\vspace{-0.5em}
\caption{Detailed ablation results for probe selection stages.}
\vspace{-0.5em}
\label{tab:probe-selection}
\centering
\scriptsize
\setlength{\tabcolsep}{3.0pt}
\begin{tabular}{l@{\quad}cccc@{\qquad}cccc@{\qquad}cccc}
\toprule
\multirow{2}{*}{\textbf{Selection stage}}
& \multicolumn{4}{c}{\textbf{Pretrained sources}}
& \multicolumn{4}{c}{\textbf{Instruct sources}}
& \multicolumn{4}{c}{\textbf{All sources}}
\\
\cmidrule(lr){2-5}
\cmidrule(lr){6-9}
\cmidrule(lr){10-13}
& AUC $\uparrow$ & pAUC $\uparrow$ & TPR $\uparrow$ & $d' \uparrow$
& AUC $\uparrow$ & pAUC $\uparrow$ & TPR $\uparrow$ & $d' \uparrow$
& AUC $\uparrow$ & pAUC $\uparrow$ & TPR $\uparrow$ & $d' \uparrow$
\\
\midrule
Random probes
& 0.761 & 0.675 & 0.347 & 0.953
& 0.741 & 0.616 & 0.224 & 0.821
& 0.751 & 0.614 & 0.143 & 0.875
\\
Non-gold pool
& 0.812 & 0.710 & 0.367 & 1.264
& 0.878 & 0.795 & 0.551 & 1.746
& 0.844 & 0.756 & 0.439 & 1.494
\\
Filtered pool
& 0.936 & 0.818 & 0.571 & 2.242
& 0.957 & 0.904 & 0.796 & 2.551
& 0.948 & 0.871 & 0.674 & 2.340
\\
Stemma
& \textbf{0.964} & \textbf{0.959} & \textbf{0.918} & \textbf{3.045}
& \textbf{0.970} & \textbf{0.938} & \textbf{0.857} & \textbf{2.945}
& \textbf{0.967} & \textbf{0.944} & \textbf{0.878} & \textbf{2.951}
\\
\bottomrule
\end{tabular}
\vspace{-1em}
\end{table}

\paragraph{Pool size and probe budget.}
We jointly vary the candidate pool size from 1,000 to 6,000 questions and the final probe budget from 10 to 100, while keeping all other settings fixed. Tables~\ref{tab:budget-pauc} and~\ref{tab:budget-auc} report the corresponding pAUC and AUC results, respectively. The default configuration is shown in bold. As the sweep was conducted independently from
the main experiment, the default-setting result differs slightly from the main-run result due to stochastic decoding and hardware-level nondeterminism.
\begin{table}[H]
\vspace{-0.5em}
\caption{Detailed pAUC results across candidate pool sizes and probe budgets.}
\vspace{-0.5em}
\label{tab:budget-pauc}
\centering
\scriptsize
\setlength{\tabcolsep}{3.0pt}
\begin{tabular}{l@{\quad}cccccccccc}
\toprule
\multirow{2}{*}{\textbf{Pool size}}
& \multicolumn{10}{c}{\textbf{Number of probes \(k\)}}
\\
\cmidrule(lr){2-11}
& 10 & 20 & 30 & \textbf{40} & 50 & 60 & 70 & 80 & 90 & 100
\\
\midrule
1,000
& 0.857 & 0.888 & 0.904 & 0.915 & 0.927
& 0.920 & 0.929 & 0.921 & 0.916 & 0.915
\\
2,000
& 0.881 & 0.934 & 0.930 & 0.934 & 0.932
& 0.928 & 0.932 & 0.937 & 0.937 & 0.936
\\
\textbf{3,000}
& 0.878 & 0.945 & 0.949 & \textbf{0.946} & 0.956
& 0.950 & 0.950 & 0.953 & 0.950 & 0.948
\\
4,000
& 0.903 & 0.935 & 0.940 & 0.951 & 0.946
& 0.951 & 0.950 & 0.944 & 0.945 & 0.943
\\
5,000
& 0.908 & 0.929 & 0.941 & 0.939 & 0.946
& 0.946 & 0.942 & 0.946 & 0.946 & 0.940
\\
6,000
& 0.887 & 0.900 & 0.920 & 0.946 & 0.938
& 0.942 & 0.945 & 0.944 & 0.950 & 0.949
\\
\bottomrule
\end{tabular}
\vspace{-1em}
\end{table}

\begin{table}[H]
\vspace{-0.5em}
\caption{Detailed AUC results across candidate pool sizes and probe budgets.}
\vspace{-0.5em}
\label{tab:budget-auc}
\centering
\scriptsize
\setlength{\tabcolsep}{3.0pt}
\begin{tabular}{l@{\quad}cccccccccc}
\toprule
\multirow{2}{*}{\textbf{Pool size}}
& \multicolumn{10}{c}{\textbf{Number of probes \(k\)}}
\\
\cmidrule(lr){2-11}
& 10 & 20 & 30 & \textbf{40} & 50 & 60 & 70 & 80 & 90 & 100
\\
\midrule
1,000
& 0.920 & 0.949 & 0.954 & 0.956 & 0.955
& 0.957 & 0.959 & 0.959 & 0.958 & 0.958
\\
2,000
& 0.946 & 0.958 & 0.962 & 0.960 & 0.960
& 0.961 & 0.962 & 0.962 & 0.965 & 0.965
\\
\textbf{3,000}
& 0.955 & 0.972 & 0.969 & \textbf{0.968} & 0.970
& 0.971 & 0.971 & 0.970 & 0.969 & 0.970
\\
4,000
& 0.955 & 0.970 & 0.968 & 0.970 & 0.970
& 0.970 & 0.969 & 0.967 & 0.969 & 0.968
\\
5,000
& 0.958 & 0.958 & 0.963 & 0.965 & 0.965
& 0.966 & 0.967 & 0.969 & 0.971 & 0.970
\\
6,000
& 0.945 & 0.956 & 0.959 & 0.963 & 0.963
& 0.964 & 0.964 & 0.964 & 0.966 & 0.965
\\
\bottomrule
\end{tabular}
\vspace{-1em}
\end{table}

\paragraph{Probe dataset.}
For the dataset ablation, candidate questions are drawn from the test splits of MMLU and MMLU-Pro and the validation splits of CommonsenseQA and CosmosQA, using up to 3,000 questions per dataset. Decision-interface calibration uses
100 questions from the validation splits of MMLU and MMLU-Pro and the training splits of CommonsenseQA and CosmosQA, ensuring that calibration and fingerprint construction use disjoint splits. The selection procedure and probe budget remain fixed across datasets.

\begin{table}[H]
\vspace{-0.5em}
\caption{Detailed ablation results for probe dataset choice.}
\vspace{-0.5em}
\label{tab:dataset}
\centering
\scriptsize
\setlength{\tabcolsep}{3.0pt}
\begin{tabular}{l@{\quad}cccc@{\qquad}cccc@{\qquad}cccc}
\toprule
\multirow{2}{*}{\textbf{Dataset}}
& \multicolumn{4}{c}{\textbf{Pretrained sources}}
& \multicolumn{4}{c}{\textbf{Instruct sources}}
& \multicolumn{4}{c}{\textbf{All sources}}
\\
\cmidrule(lr){2-5}
\cmidrule(lr){6-9}
\cmidrule(lr){10-13}
& AUC $\uparrow$ & pAUC $\uparrow$ & TPR $\uparrow$ & $d' \uparrow$
& AUC $\uparrow$ & pAUC $\uparrow$ & TPR $\uparrow$ & $d' \uparrow$
& AUC $\uparrow$ & pAUC $\uparrow$ & TPR $\uparrow$ & $d' \uparrow$
\\
\midrule
CommonsenseQA
& 0.915 & 0.770 & 0.490 & 1.856
& \textbf{0.977} & \textbf{0.961} & \textbf{0.898} & 2.936
& 0.948 & 0.868 & 0.714 & 2.232
\\
CosmosQA
& 0.889 & 0.783 & 0.551 & 1.748
& 0.951 & 0.874 & 0.694 & 2.601
& 0.925 & 0.815 & 0.582 & 2.031
\\
MMLU-Pro
& 0.947 & 0.938 & 0.878 & 2.736
& 0.948 & 0.916 & 0.837 & 2.673
& 0.948 & 0.929 & 0.857 & 2.635
\\
MMLU (default)
& \textbf{0.964} & \textbf{0.959} & \textbf{0.918} & \textbf{3.045}
& 0.970 & 0.938 & 0.857 & \textbf{2.945}
& \textbf{0.967} & \textbf{0.944} & \textbf{0.878} & \textbf{2.951}
\\
\bottomrule
\end{tabular}
\vspace{-1em}
\end{table}

\paragraph{Background model.}
For the background-model ablation, we compare
\texttt{Phi-3.5-mini-instruct}, \texttt{Yi-1.5-9B-Chat}, and \texttt{Gemma-2-9B-IT}, while keeping all other settings fixed. In the combined setting, we pool the permutation-level decisions from all three models and compute background alignment as the fraction matching the source decision.

\begin{table}[H]
\vspace{-0.5em}
\caption{Detailed ablation results for background model choice.}
\vspace{-0.5em}
\label{tab:background}
\centering
\scriptsize
\setlength{\tabcolsep}{3.0pt}
\begin{tabular}{l@{\quad}cccc@{\qquad}cccc@{\qquad}cccc}
\toprule
\multirow{2}{*}{\textbf{Background model}}
& \multicolumn{4}{c}{\textbf{Pretrained sources}}
& \multicolumn{4}{c}{\textbf{Instruct sources}}
& \multicolumn{4}{c}{\textbf{All sources}}
\\
\cmidrule(lr){2-5}
\cmidrule(lr){6-9}
\cmidrule(lr){10-13}
& AUC $\uparrow$ & pAUC $\uparrow$ & TPR $\uparrow$ & $d' \uparrow$
& AUC $\uparrow$ & pAUC $\uparrow$ & TPR $\uparrow$ & $d' \uparrow$
& AUC $\uparrow$ & pAUC $\uparrow$ & TPR $\uparrow$ & $d' \uparrow$
\\
\midrule
Phi-3.5 (default)
& 0.964 & \textbf{0.959} & \textbf{0.918} & \textbf{3.045}
& 0.970 & 0.938 & 0.857 & 2.945
& 0.967 & 0.944 & 0.878 & \textbf{2.951}
\\
Yi-1.5
& 0.958 & 0.940 & 0.857 & 2.531
& 0.964 & 0.955 & 0.898 & 2.853
& 0.959 & 0.939 & 0.857 & 2.628
\\
Gemma-2
& 0.955 & 0.959 & \textbf{0.918} & 2.654
& 0.968 & 0.942 & 0.878 & 2.901
& 0.960 & 0.946 & 0.878 & 2.700
\\
Combined
& \textbf{0.966} & 0.957 & \textbf{0.918} & 2.983
& \textbf{0.979} & \textbf{0.958} & \textbf{0.918} & \textbf{3.077}
& \textbf{0.973} & \textbf{0.952} & \textbf{0.898} & 2.950
\\
\bottomrule
\end{tabular}
\vspace{-1em}
\end{table}
\section{Runtime Measurement}
\label{app:runtime}

\paragraph{Measurement protocol.}
We measure the wall-clock runtime of all methods on the same NVIDIA L40S GPU with 48\,GB of memory. Fingerprint construction is measured per source model and reported as the median over the 14 source checkpoints. Fingerprint verification measures the suspect-side computation required to evaluate one source--suspect pair and is reported as the median over the corresponding verification executions, assuming that the source fingerprint has already been constructed. The measurements cover the principal computational procedures, such as model inference and, where applicable, prompt optimisation involving backward passes, but exclude model loading. Reported runtimes are rounded to the nearest second.

\paragraph{LLMmap.}
The reported construction time includes evaluating each source under 100 prompt configurations on the 8 diagnostic queries, encoding the resulting responses, and averaging the feature vectors to construct the gallery representation. The reported verification time includes evaluating each suspect once on the same 8 queries and encoding its responses using the pretrained feature extractor.

\paragraph{LLMPrint.}
The reported construction time includes optimising one adversarial suffix for each of 100 word pairs, each consisting of two single-token words, over 200 GCG steps, and estimating the source model's preference direction for each pair using 100 repeated one-token generations. The reported verification time includes querying each suspect model with the fixed optimised suffixes, estimating its preference direction for each pair using 100 repeated one-token generations, and comparing these directions with the stored source preferences.

\paragraph{MPT.}
The reported construction time includes querying each source on 5,000 prompts with single-token generation. The reported verification time includes querying each suspect once on the same prompt set and matching the resulting continuations against the cached source continuations to compute the hit rate.

\paragraph{ZeroPrint.}
The reported construction time includes querying each source on 2 HumanEval base prompts and 4 word-substitution variants per prompt, with 20 generations for each query, embedding the responses, and constructing the Jacobian-based fingerprint. The reported verification time includes applying the same procedure to each suspect.

\paragraph{Stemma.}
The reported construction time includes calibrating the source model and the configured background model on 100 held-out validation questions, followed by evaluating them on 3,000 candidate questions under cyclic option permutations to obtain the source-decision and background-alignment statistics used for probe selection. The reported verification time includes calibrating each suspect on the same held-out set and querying it on the 40 source-specific probes under four cyclic permutations.
\section{The Stemmatic Analogy Behind Stemma}
\label{app:naming}

\paragraph{Stemmatic analysis.}
Manuscript studies draw on several complementary approaches, including palaeography, codicology, and stemmatics. Palaeography examines handwriting and scribal practices, while codicology examines the material construction and production of manuscripts. Stemmatics instead focuses on reconstructing the transmission history of works preserved in multiple manuscript copies, particularly when the original text and many intermediate copies no longer survive and only a set of extant manuscripts, commonly termed
\emph{witnesses}, remains \citep{heikkilaThematicSectionStudia2016}.

To infer the genealogy of the transmitted text, stemmatic analysis aligns witnesses at corresponding textual locations and compares their variant readings, including substitutions, omissions, additions, and transpositions. It therefore relies on structured patterns of textual variation rather than treating overall visual or stylistic resemblance as the primary genealogical signal. The genealogical value of a shared variant depends on how readily it could have arisen independently. Variants that are widespread across the textual tradition or easily reproduced provide limited evidence, whereas distinctive variants are more informative because their preservation by multiple witnesses may indicate descent from a common exemplar. Stemmatic inference therefore considers not merely the presence of shared variation, but its genealogical informativeness \citep{andrewsAnalysisVariationSignificance2016}. The inferred relationships are conventionally represented by a \emph{stemma codicum}, or \emph{stemma}, a tree-like representation of textual descent among surviving witnesses and reconstructed intermediate exemplars.

\paragraph{Analogy to model provenance.}
The name Stemma reflects this methodological analogy. Candidate questions correspond to aligned textual locations, model instances to manuscript witnesses, and induced decisions to variant readings. Existing response-based fingerprints derive provenance signals from observable characteristics of generated responses, much as palaeography and codicology draw on handwriting and material features, whereas Stemma maps model behaviour into a common finite decision space and compares structured decision variants across aligned probes. Decisions readily reproduced by unrelated models provide limited provenance evidence, just as common or independently arising readings provide limited evidence of textual affiliation. By contrast, stable, robust, and specific decisions preserved by a suspect resemble distinctive inherited variants. Stemma therefore tests model provenance through structured patterns of induced decision region inheritance rather than through response-level characteristics tied to particular surface realisations.

\end{document}